\documentclass[aps,pra,reprint,showpacs]{revtex4-1}

\usepackage{graphicx}
\usepackage{dcolumn}
\usepackage{bm}
\usepackage{color}
\usepackage{scrextend}		
\usepackage{url}
\usepackage{epstopdf}
\usepackage{hyperref}

\begin{document}

\title{Structural and Magnetic Properties of Pt/Co/Mn-Based Multilayers}

\author{Martin Lonsky}
\email{Electronic mail: mlonsky@physik.uni-frankfurt.de}
\thanks{Present address: Institute of Physics, Goethe University Frankfurt, 60438 Frankfurt, Germany}
\author{Myoung-Woo Yoo}
\author{Yi-Siou Huang}
\author{Jiangchao Qian}
\author{Jian-Min Zuo}
\author{Axel Hoffmann}
\affiliation{Materials Research Laboratory and Department of Materials Science and Engineering, University of Illinois at Urbana-Champaign, Urbana, Illinois 61801, USA}

\date{\today}

\begin{abstract}
Magnetic multilayers are a rich class of materials systems with numerous highly tunable physical parameters that determine both their magnetic and electronic properties. Here we present a comprehensive experimental study of a novel system, Pt/Co/Mn, which extends the group of Pt/Co/X ($\mathrm{X}=$ metal) multilayers that have been investigated thus far. We demonstrate that an increasing Co layer thickness changes the magnetic anisotropy from out-of-plane to in-plane, whereas the deposition of thicker Mn layers leads to a decrease in the saturation magnetization. Temperature-dependent magnetometry measurements reinforce the hypothesis of antiferromagnetic coupling at the Co/Mn interfaces being responsible for the observed Mn thickness dependence of the magnetization reversal. Moreover, magneto-optical imaging experiments indicate systematic changes in magnetic domain patterns as a function of the Co and Mn layer thickness, suggesting the existence of bubble-like domains -- potentially even magnetic skyrmions -- in the case of sufficiently thick Mn layers, which are expected to contribute to a sizeable Dzyaloshinskii-Moriya interaction in the multilayer stacks. We identify Pt/Co/Mn as a highly complex multilayer system with strong potential for further fundamental studies and possible applications. 
\end{abstract}

\maketitle

\section{Introduction} \label{intro}
Metallic magnetic films and multilayers are materials systems that offer a high degree of tunability in regards to their magnetic and electronic properties \cite{Schuller_1999, Bennett_1994, Camley_1993}, and are key to contemporary magnetic applications based on the concepts of spintronics and magnonics \cite{Vedmedenko_2020, Back2020}. The space of controllable parameters includes, but is not limited to, the type of chemical elements, the layer thickness, the number of layer repetitions, or the choice of the deposition method. Especially for magnetic multilayers there is a wide variety of novel physical phenomena that emerge at the interfaces \cite{Hellman_2017}. One particularly interesting phenomenon in such multilayers is given by the existence of an antisymmetric exchange interaction, the Dzyaloshinskii-Moriya interaction (DMI) \cite{Dzyaloshinskii1958, Moriya1960}, which under certain conditions may give rise to the presence of magnetic skyrmions even at room temperature \cite{Jiang2017}. In condensed matter physics, magnetic skyrmions are topologically distinct chiral spin textures which may find use in future information-processing devices or non-volatile memory solutions. Such devices may for instance be realized by creating, moving, and reading skyrmions in a fast and reliable manner \cite{Woo2016, Woo_2018, Everschor-Sitte2018}, or by harnessing the coupled internal eigenmodes of chiral magnetic textures \cite{Kim2017, Kim2018, Lonsky2020a, Lonsky2020, Chen_2021}.

One of the most promising classes of multilayers is given by Pt/Co/X systems, where X corresponds to an additional metallic layer that is required to break spatial inversion symmetry and thus contributes to a significant DMI \cite{Jia2020, Ajejas2021} -- we note that even in pure Pt/Co multilayer systems a weak DMI has been reported due to the different character of Pt/Co and Co/Pt interfaces \cite{Wells2017}. Pt/Co/X systems are based on scientifically well-understood Pt/Co multilayers which have been studied extensively for several decades. Under certain conditions, the addition of a third layer to the Pt/Co-bilayer system may lead to the presence or enhancement of a DMI. In this regard, Ajejas \textit{et al.}\ recently reported on the dependence of the interfacial DMI strength on the work function difference at the Co/X interface, whereby the authors have considered various metallic materials X, such as Pd, Ru, and Al \cite{Ajejas2021}. Further works have also focused on several other Pt/Co/X-based multilayers, including Co/Ni/Pt \cite{Brock2020}, Pt/Co/W \cite{Jiang2019}, Pt/Co/Ta \cite{Woo2016}, Pt/Co/Ir \cite{Khadka2018}, Pt/Co/Ho \cite{Liu_2020}, and Pt/Co/Cu \cite{Schlotter2018}. In this context, the use of Mn as the third metallic layer X in Pt/Co/X has not been reported thus far, even though the significantly different work functions $\Phi$ of Co and Mn, $\Phi_{\mathrm{Co}}=5.0$\,eV and $\Phi_{\mathrm{Mn}}=4.1$\,eV \cite{Michaelson1977}, suggest the presence of a considerable interfacial DMI. In addition, further interesting and complex physical phenomena can be expected in Pt/Co/Mn systems due to possible ferro- or antiferromagnetic interlayer and interfacial coupling between Co and Mn, as well as due to the formation of different Mn phases and the occurence of intermixing at the Co/Mn interface, as reported by various articles on Co/Mn multilayers \cite{Kai_2000, Zhang_2014, Henry_1996, Ishida_2004, Uchiyama_1996, Ishida_2003}. Despite being well-established in the literature, interdiffusion is an often overlooked -- but nevertheless important -- phenomenon that can occur even in room temperature-deposited films and typically determines the properties of magnetic heterostructures, see for example reports on Co/Gd/Pt \cite{Nishimura_2021, Nishimura2020} or Co/Pt multilayers \cite{Su_2009, Bandiera_2012}. 

Motivated by the prospect of rich and complex physical phenomena in a novel metallic heterostructure system, we fabricated polycrystalline [Pt/Co/Mn]$_n$ multilayer samples by means of direct current (dc) magnetron sputtering and carried out a comprehensive characterization of their structural and magnetic properties. In the remainder of this article, we will demonstrate that the magnetic characteristics of Pt/Co/Mn-based multilayers are highly sensitive to the Co and Mn layer thicknesses and can be tuned systematically by careful execution of the deposition processes. Furthermore, systematic changes in the magnetic domain structure as revealed by magneto-optical imaging experiments will be discussed in light of the possible emergence of a DMI and other magnetic interactions.
             
\section{Experimental Methods}

Magnetic Ta(2.0)/[Pt(1.5)/Co(x)/Mn(y)]$_n$/Pt($t_{\mathrm{cap}}$) multilayers have been deposited at room temperature on thermally oxidized Si(100) substrates by means of dc and pulsed dc magnetron sputtering using an AJA International ATC 2200 system. Numbers in parentheses indicate the layer thickness in nm. Furthermore, $x$ and $y$ correspond to the thicknesses of the Co and Mn layers, respectively, and the Pt cap layer thickness $t_{\mathrm{cap}}$ was chosen to be either $1.5$ or $2.0\,$nm for all samples. The sputtering power has been set to $30\,$W for all materials and the vacuum chamber was kept at a base pressure better than $2\times 10^{-8}\,$Torr. During deposition the argon pressure was 3 mTorr. The combination of relatively small power values and a low partial Ar pressure was chosen to increase the deposition time and thereby reduce the error in the thickness of the ultrathin layers. Subsequently, structural characterization has been performed by x-ray reflectometry (XRR) on a Bruker D8 Advance system. Modeling and fitting was carried out with the Bruker LEPTOS software package. 
Scanning transmission electron microscope (STEM) studies of a selected multilayer stack complement the structural analysis of the Pt/Co/Mn heterostructures. Towards this end, STEM samples were prepared using a focused ion beam (FIB; Thermo Scios2 Dual Beam) microscope lift-out milling technique and the sample was thinned to $\sim 50\,$nm with a careful cleaning processing under $5\,$kV and $2\,$kV. To investigate the elemental configuration, energy dispersive x-ray spectroscopy (EDS) has been operated on a STEM (FEI Themis Z Advanced Probe Aberration Corrected Analytical TEM/STEM) at an accelerating voltage of $300\,$kV (Schottky X-FEG gun) and equipped with 4-crystal EDS systems (FEI Super-X) in STEM mode with a semi-angle of $18\,$mrad. All STEM-EDS data were collected for more than $30\,$min with a $10\, \mathrm{\mu}$s dwell time and $\sim 100\,$pA probe current.
Magnetic properties have been measured in a Magnetic Properties Measurement System (MPMS) by Quantum Design, using dc superconducting quantum interference device (SQUID) magnetometry as well as the vibrating sample magnetometer (VSM) option, both yielding consistent results. For samples with a reduced magnetic moment, we have observed the presence of considerable background signals, presumably related to paramagnetic and diamagnetic contributions of the silicon substrate or the sample holder. Therefore, we have also recorded hysteresis loops for selected samples using a polar magneto-optical Kerr effect (MOKE) magnetometry setup, where no spurious signals occured -- see Appendix for details. Furthermore, we have utilized a home-built Kerr microscope for magneto-optical imaging of magnetic domain structures in the Pt/Co/Mn samples. 
  
\section{Results and Discussion} \label{RESULTS}
\subsection{Structural Properties}

\begin{figure}
\centering
\includegraphics[width=8.5 cm]{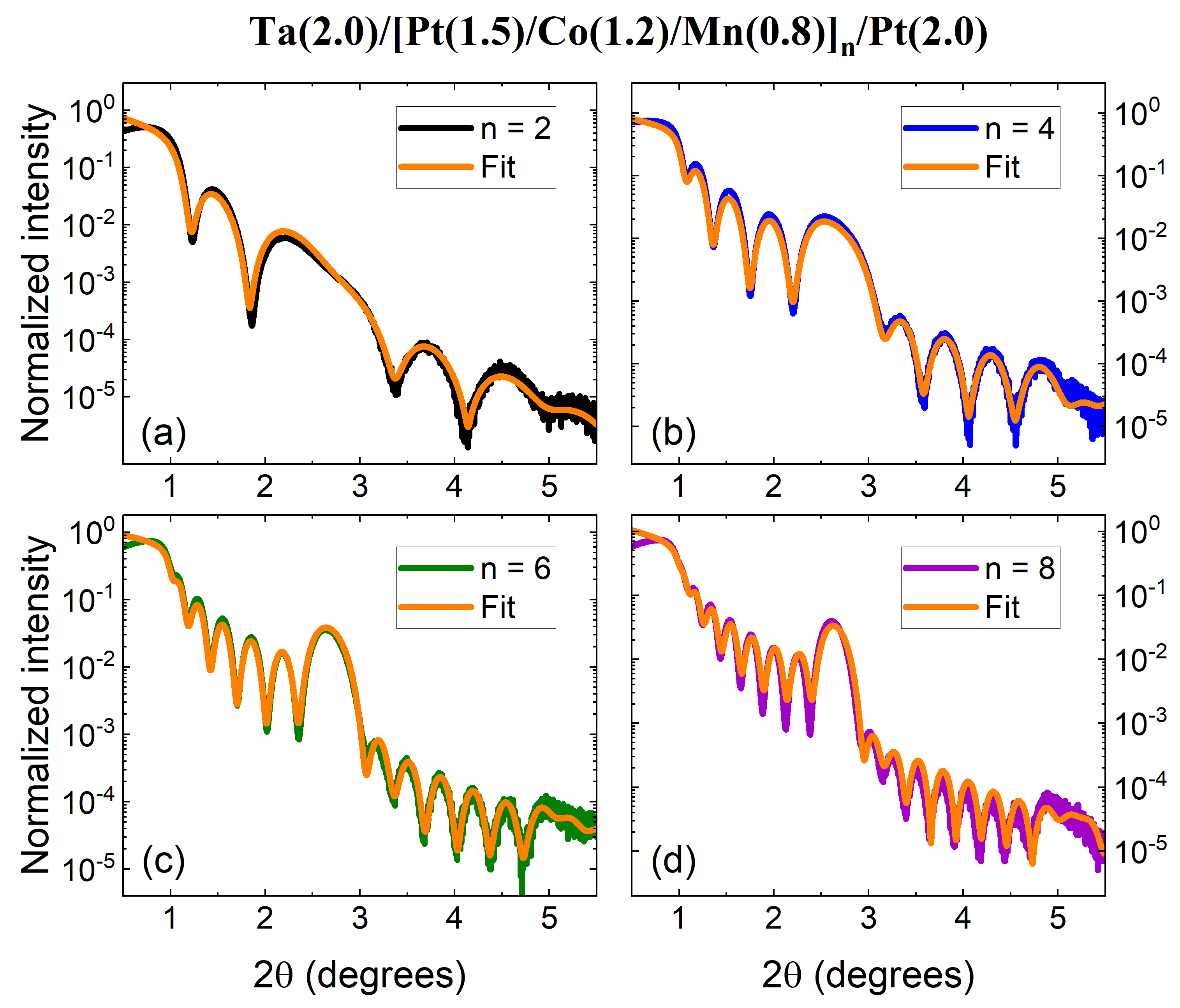}
\caption{XRR spectra for samples with varying number of repetitions $n=2,4,6,8$. Depicted are the experimental data as well as fits (orange curves) for angles $2\theta$ ($\theta$: incident angle) ranging from $0.5$ to $5.5$ degrees. The maximum interfacial roughness is around $0.4$\,nm for all samples, see Table \ref{tab:table1}.} 
\label{XRR_NDEP}%
\end{figure}

In a first step, we have carried out an extensive structural characterization with XRR in order to demonstrate the accuracy and reliability of the magnetron sputtering of [Pt/Co/Mn]$_n$. Figure \ref{XRR_NDEP} shows the XRR spectra for angles $2\theta$ between $0.5$ and $5.5$ degrees for a set of four samples with a different number $n$ of Pt/Co/Mn trilayer repetitions. Here, $\theta$ corresponds to the incident angle of the x-ray beam. Clearly, Kiessig fringes with different wavelengths can be observed for the respective samples. To obtain a better understanding of the data, theoretical XRR spectra are modeled and then fitted to the experimental data by using the LEPTOS software suite. An excellent agreement between the experimental data and the fits (orange curves) can be seen for all four samples. The utilized fit parameters are presented in Table \ref{tab:table1}. Here, $t_{\mathrm{M}}$ corresponds to the layer thickness of material M, and $\sigma_{\mathrm{M}}$ indicates the layer roughness. It is plausible to assume that the individual Pt, Co, and Mn layer thickness and roughness values are identical for every repetition of the Pt/Co/Mn trilayer. In the bottom row, the nominal thickness values $t_{\mathrm{M}}$ are also presented for the sake of clarity. As can be seen in the table, the maximum interfacial roughness $\sigma$ for all multilayer specimens has been determined to be lower than $0.4\,$nm. This implies that the sputtered films are of good quality.

\begin{table*}
\caption{\label{tab:table1}Overview of utilized fit parameters for XRR spectra of Ta(2.0)/[Pt(1.5)/Co(1.2)/Mn(0.8)]$_n$/Pt(2.0) samples with $n=2,4,6,8$ Pt/Co/Mn trilayer repetitions as displayed in Fig.\ \ref{XRR_NDEP}. $t_{\mathrm{M}}$ indicates the layer thickness of material M, and $\sigma_{\mathrm{M}}$ the corresponding layer roughness. Bottom row indicates the nominal thickness values $t_{\mathrm{M}}$.} 
\begin{ruledtabular}
\begin{tabular}{ccccccccccc}
$n$&$t_{\mathrm{Ta}}$ [nm]&$\sigma_{\mathrm{Ta}}$ [nm]&$t_{\mathrm{Pt}}$ [nm]&$\sigma_{\mathrm{Pt}}$ [nm]&$t_{\mathrm{Co}}$ [nm]&$\sigma_{\mathrm{Co}}$ [nm]&$
t_{\mathrm{Mn}}$ [nm]&$\sigma_{\mathrm{Mn}}$ [nm]&$t_{\mathrm{Pt,cap}}$ [nm]&$
\sigma_{\mathrm{Pt,cap}}$ [nm]\\
\hline
$2$ & $1.98$ & $0.18$ & $1.46$ & $0.37$ & $1.18$ & $0.20$& $0.88$ & $0.24$ & $2.00$ & $0.29$\\
$4$ & $1.98$ & $0.22$ & $1.50$ & $0.27$ & $1.09$ & $0.29$& $0.86$ & $0.34$ & $2.12$ & $0.28$\\
$6$ & $1.96$ & $0.31$ & $1.54$ & $0.29$ & $1.17$ & $0.21$& $0.73$ & $0.34$ & $2.19$ & $0.38$\\
$8$ & $2.01$ & $0.28$ & $1.61$ & $0.26$ & $1.08$ & $0.20$& $0.84$ & $0.32$ & $2.28$ & $0.29$\\
\hline
Nominal $t_{\mathrm{M}}$ & $2.00$ &  & $1.50$ & & $1.20$ & & $0.80$ &  & $2.00$ & \\
\end{tabular}
\end{ruledtabular}
\end{table*}

\begin{figure}
\centering
\includegraphics[width=8.5 cm]{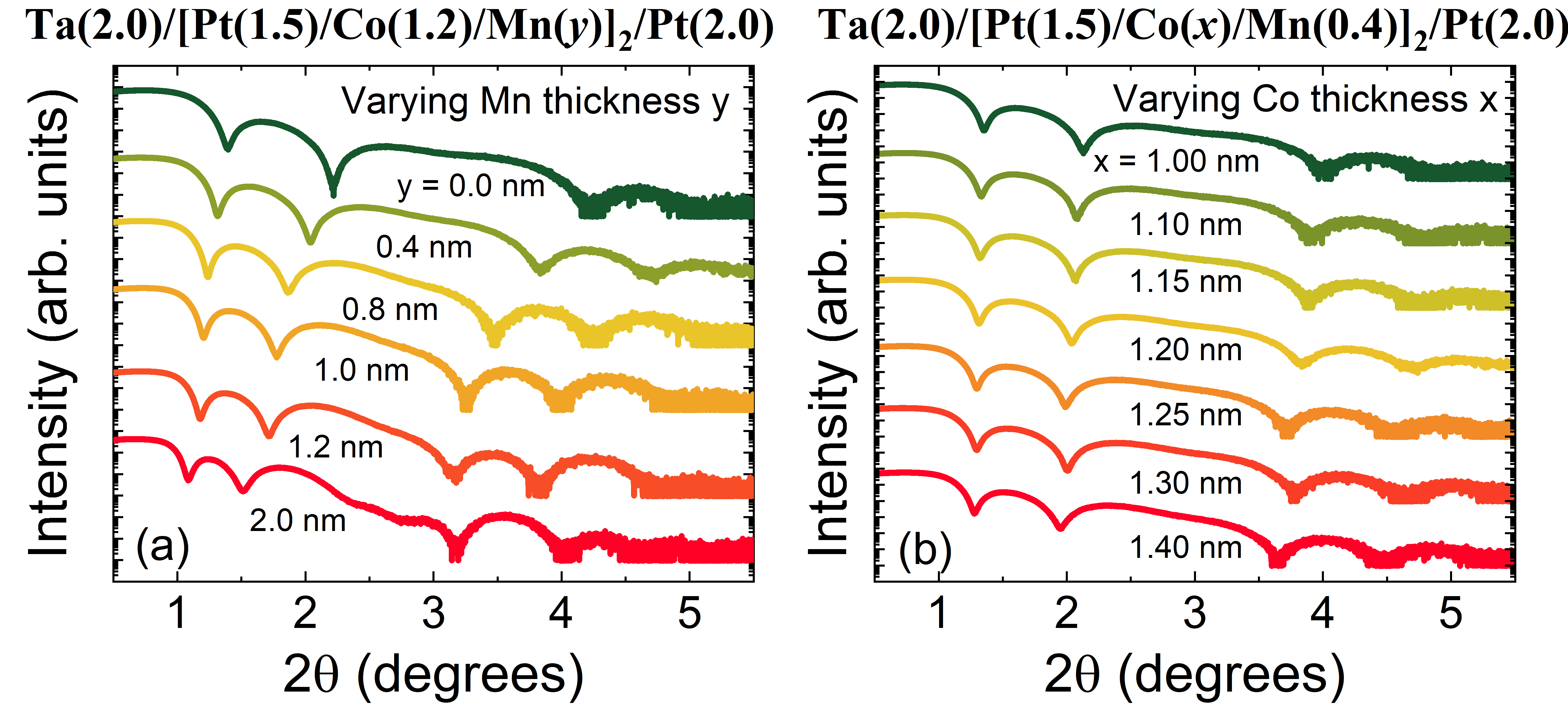}
\caption{XRR spectra for samples with varying (a) Mn thickness and (b) Co thickness. Curves have been shifted along the $y$-axis for the sake of clarity. The systematic behavior clearly demonstrates the accuracy and reliability of the established magnetron sputtering procedures.} 
\label{XRR_COMNDEP}%
\end{figure}

In Fig.\ \ref{XRR_COMNDEP}, we present XRR spectra for various samples with $n=2$ repetitions and a varying Mn or Co layer thickness. All spectra are shifted with respect to each other along the $y$-axis for the sake of clarity. We observe systematic changes in the spectra in accordance with the respective thickness variations, which are also reflected in the corresponding fits (not shown). While these data underline the high quality of the sputter-deposited Pt/Co/Mn heterostructures, we point out that the analysis of XRR spectra is complicated by the complexity of the investigated multilayers, similar atomic numbers of Mn and Co, as well as possible intermixing of these elements. To gain a deeper understanding of the structural properties in general, we also performed STEM investigations on a selected [Pt/Co/Mn]$_{n}$ sample with $n=10$. 

\begin{figure}
\centering
\includegraphics[width=8.5 cm]{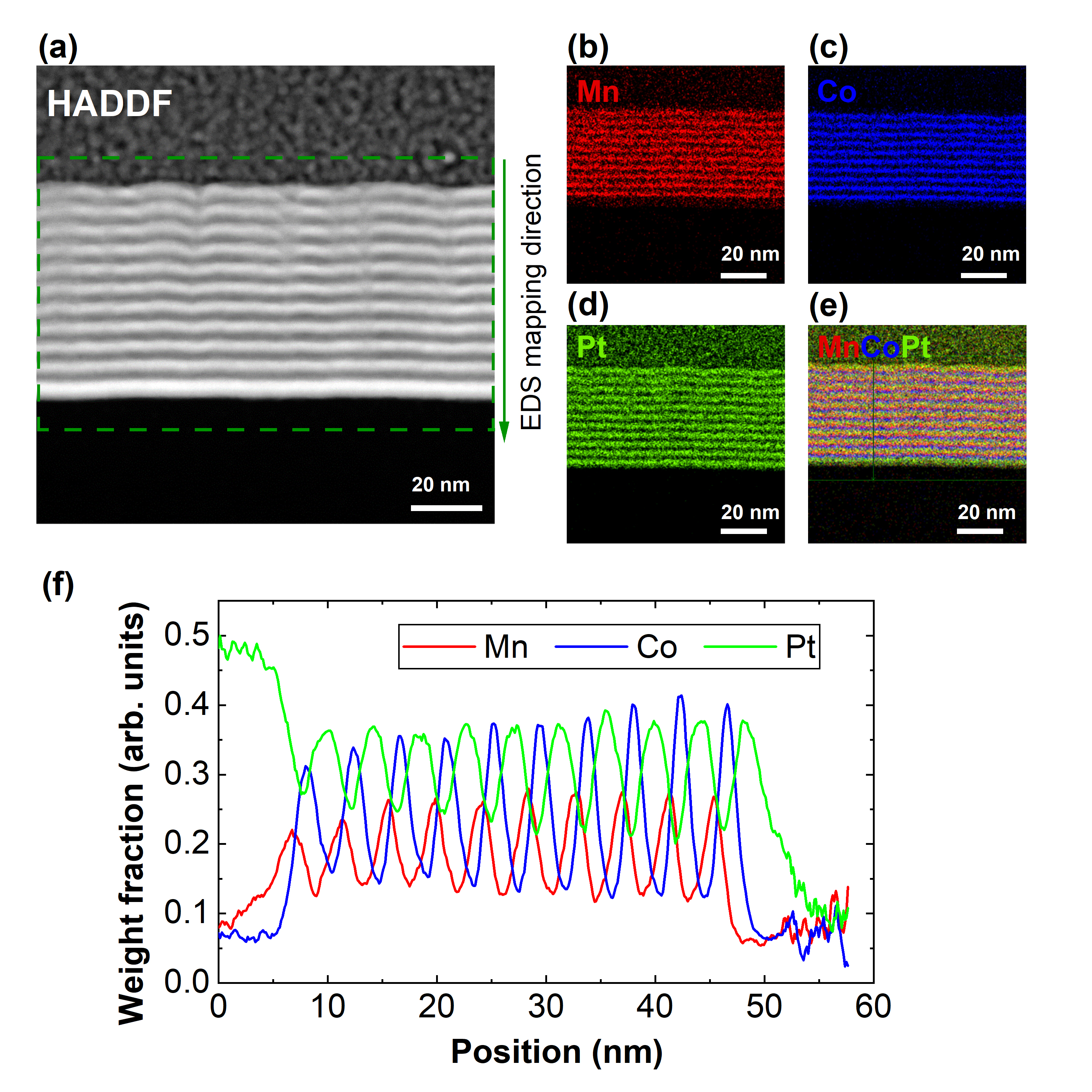}
\caption{(a) High-angle annular dark-field imaging (HAADF) scanning transmission electron microscopy (STEM) micrograph of selected sample, Ta(2.0)/[Pt(1.5)/Co(1.2)/Mn(1.2)]$_{10}$/Pt(1.5). EDS elemental mapping images are shown for (b) Mn, (c) Co, (d) Pt individually, as well as for (e) all three elements. (f) Weight fraction mapping of Mn, Co and Pt across the multilayer stack as determined from STEM-EDS measurements. An adjacent averaging filter of 5 pixels is used for smoothing the composition area profiles in (f).} 
\label{TEM_IMG}%
\end{figure}

STEM results are shown in Fig.\ \ref{TEM_IMG}. Figure \ref{TEM_IMG}(a) contains an STEM image of a selected Ta(2.0)/[Pt(1.5)/Co(1.2)/Mn(1.2)]$_{10}$/Pt(1.5) sample acquired in high-angle annular dark-field imaging (HADDF) mode. The multilayer stack is clearly visible and covered with a Pt protection layer deposited during the FIB process in which the STEM sample was prepared. 
The four pictures in the top right corner correspond to EDS elemental mapping images for the case of (b) Mn, (c) Co, (d) Pt as well as (e) the combination of these chemical elements. 
Panel (f) illustrates the weight fraction mapping of Mn, Co and Pt across the multilayer stack. The EDS mapping area and direction ($0$ to $60\,$nm) are indicated in Fig.\ \ref{TEM_IMG}(a) by the green box and arrow, respectively. The distance between two adjacent peaks in panel (f) has been determined as $4.2\pm 0.3\,$nm for Pt, while it is $4.3\pm 0.2\,$nm for both Co and Mn. This is in reasonable agreement with the expected values of $3.9\,$nm for the given sample.
The existence of ten well-defined maxima for each of the three elements demonstrates that the individual layers are separated and only weak intermixing effects may have occured at the interfaces. In summary, the XRR and STEM-EDS data verify the precise thickness control in our sputter deposition procedures for Pt/Co/Mn-based multilayers and thereby substantiate the analysis of magnetic properties that is presented hereinafter. 

\subsection{Magnetic Hysteresis Loops at Room Temperature}

As a first step, we have recorded room-temperature magnetic hysteresis curves with SQUID magnetometry measurements. Because of the non-trivial background signal that occurs due to the para- and diamagnetic contributions of the sample holder, possible impurities and the Si substrate, we also measured hysteresis curves of selected samples by means of a MOKE setup which are in excellent agreement with the SQUID measurements, see Appendix for a detailed discussion. 

\begin{figure}
\centering
\includegraphics[width=8.5 cm]{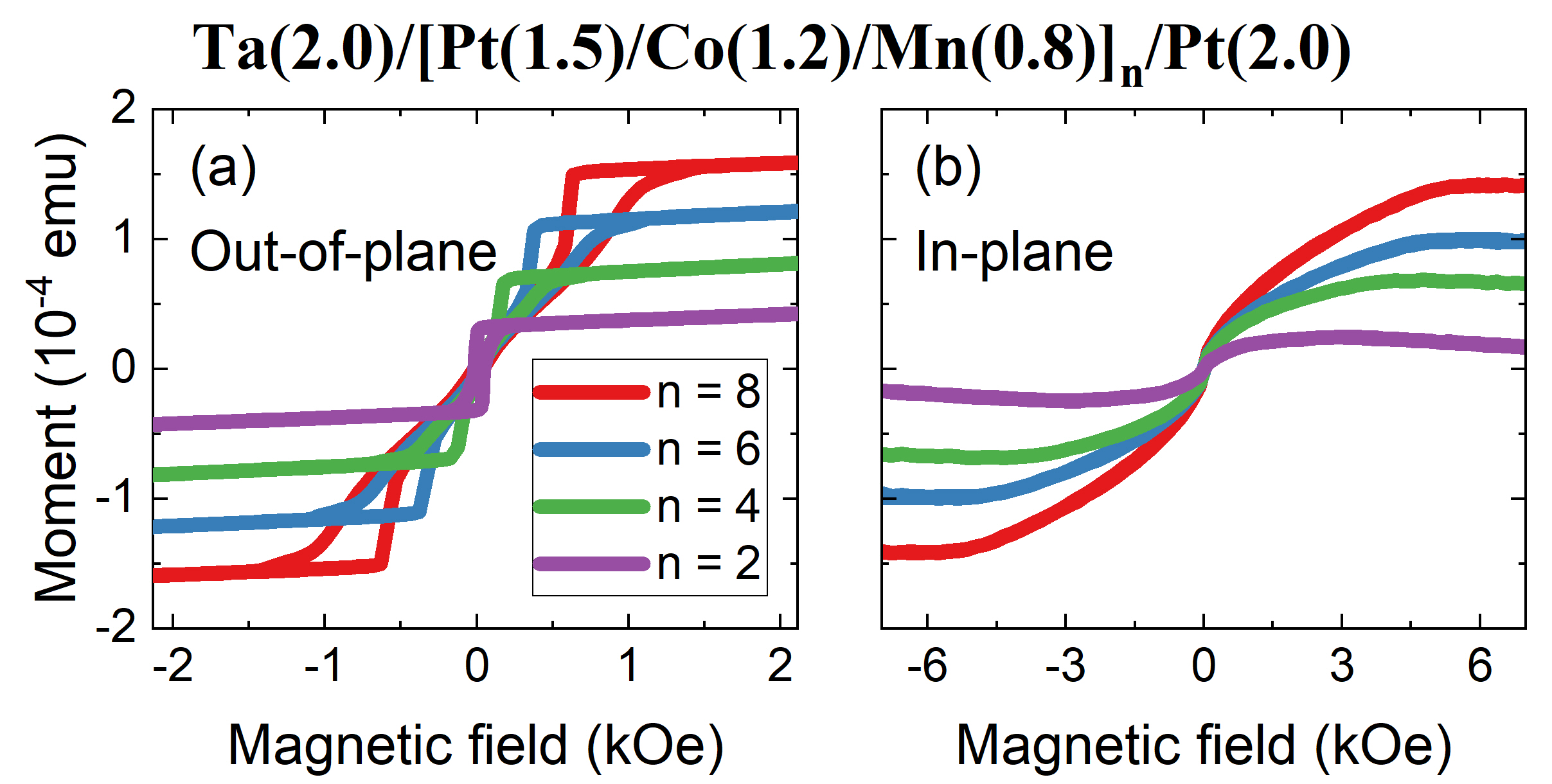}
\caption{Magnetic hysteresis loops at room temperature for [Pt/Co/Mn]$_n$ with varying $n$ and the external magnetic field applied (a) out of the film plane and (b) in the film plane. Systematic changes in the hysteresis curves can be explained by modified magnetic energy terms for varying $n$.} 
\label{OOPIP}%
\end{figure}

In the following, we consider samples with a different number $n$ of Pt/Co/Mn trilayer repetitions. Experiments have been performed at $300$\,K and the external magnetic field has been applied either perpendicular to the film plane or parallel. Figure \ref{OOPIP} (a) displays hysteresis loops (field perpendicular to film plane) for the four different samples on which the XRR analysis has been carried out, see Fig.\ \ref{XRR_NDEP}. As it is intuitively expected, due to the higher magnetic (Co) volume, the saturation magnetic moment becomes larger for samples with a higher number of layers. In general, the increasing number of repetitions can lead to modifications in various magnetic energy terms such as the anisotropy or dipolar interaction \cite{Wang_2020}, which explains the observed systematic changes in the hysteresis curves. Furthermore, the sheared nature of OOP hysteresis loops is common for systems that exhibit labyrinth domains which may transform into skyrmions, see for example Refs.\ \cite{Lin_2018, Soumyanarayanan_2017}. In addition, when comparing the curves to the ones in Figure \ref{OOPIP} (b), for which an in-plane magnetic field was applied, it is evident that the samples exhibit a perpendicular magnetic anisotropy. More specifically, the in-plane measurement curves show a much weaker hysteresis than their out-of-plane counterparts.

\begin{figure}
\centering
\includegraphics[width=6.5 cm]{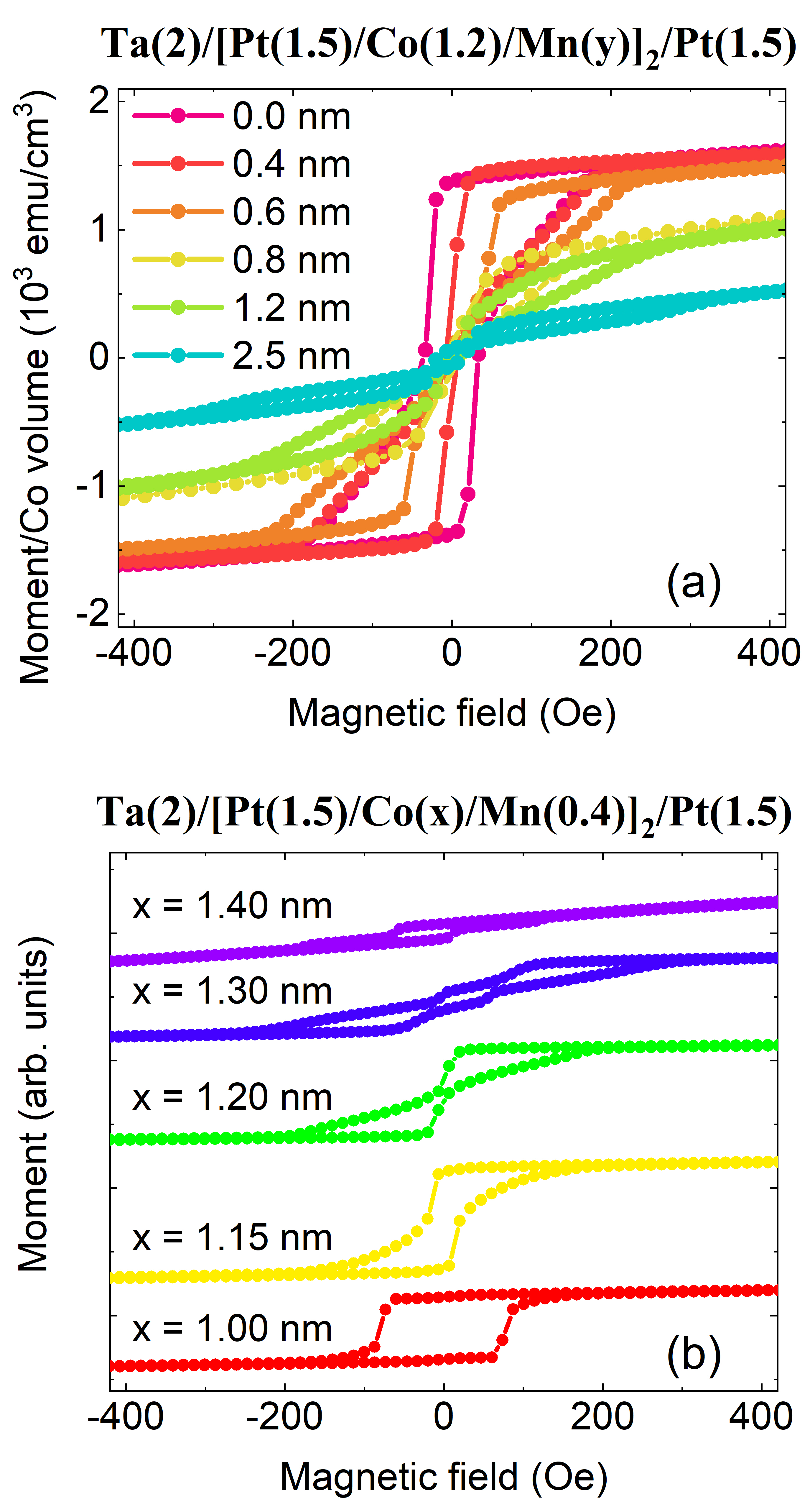}
\caption{Magnetic hysteresis curves for [Pt/Co/Mn]$_2$ at $300\,$K for varying (a) Mn thickness $y$ and (b) Co thickness $x$. Decreasing $M_\mathrm{s}$ for higher $y$ is related to emergent antiferromagnetic coupling within the multilayer stacks. Interfacial anisotropy is strongly dependent on $x$ and thereby explains the change from OOP to IP anisotropy toward higher $x$.} 
\label{HYSTTHICK}%
\end{figure}

Next, we consider the hysteresis loops for samples with varying Mn and Co thickness. Hysteresis curves for Ta(2)/[Pt(1.5)/Co(1.2)/Mn($y$)]$_2$/Pt(1.5) with varying Mn thickness values $y$ are displayed in Fig.\ \ref{HYSTTHICK}(a), whereby the total magnetic moment divided by the volume of the Co layers is plotted on the ordinate. Most importantly, aside from minor changes in the shape of the hysteresis loops, the saturation magnetization $M_{\mathrm{s}}$ strongly decreases as a function of increasing Mn thickness $y$. Due to the non-trivial background signal occurring in SQUID measurements of samples with low magnetic moment, we have verified the trend of a decreasing $M_{\mathrm{s}}$ with laser-based polar MOKE measurements, which show the Kerr rotation signal to be reduced as a function of increasing Mn thickness (not shown). 
The unusual dependence of the magnetic properties on the Mn layer thickness can be explained by an interfacial exchange coupling between adjacent Co and Mn layers, which may even be accompanied by an antiferromagnetic ordering in the Mn layers in analogy to the findings by Henry and Ounadjela in epitaxial Co/Mn multilayers \cite{Henry_1996}. Furthermore, in the aforementioned work it is discussed that at the same time an interlayer exchange coupling between adjacent Co layers through the Mn spacer may occur. However, we do not find evidence for the latter scenario due to the lack of an exchange bias effect or changes in the shape of hysteresis loops as a function of Pt/Co/Mn repetition number $n$. Apart from this, we cannot rule out slight interdiffusion between the Co and Mn layers, in analogy to the intermixing of Co and Pt reported in Ref.\ \cite{Bandiera_2012}. The formation of a Co-Mn alloy at the interface may also lead to the presence of antiferromagnetic coupling within the layers, see for instance Ref.\ \cite{Menshikov1985}. In detail, Men'shikov \textit{et al.}\ have demonstrated that Co-Mn alloys exhibit antiferromagnetism or ferromagnetism, or even a mixture of both, depending on the exact composition. Furthermore, at low temperature such alloys may display a behavior similar to a spin glass \cite{Menshikov1985}. We also point out that for low nominal Mn thickness values in some of our samples, there is the possibility that Mn layers have not been deposited in a perfectly homogeneous way and instead remained in the island-growth regime which is common for sputtered ultrathin films \cite{Waring_2020} and may affect the magnetic properties.
Before a more in-depth investigation of the complex Mn thickness dependence, we now turn to the magnetic properties of Ta(2)/[Pt(1.5)/Co($x$)/Mn(0.4)]$_2$/Pt(1.5) with varying Co layer thickness $x$. The curves in Fig.\ \ref{HYSTTHICK}(b) have been shifted with respect to each other for the sake of clarity. Clearly, the magnetization reversal changes from a relatively strong OOP anisotropy behavior toward a more IP-like systematics for increasing Co layer thicknesses, where hysteresis curves become more sheared. This is in good agreement with previous works on Co/Pt multilayers, where the origin of perpendicular magnetic anisotropy was proven to be interfacial anisotropy, which is highly sensitive on the Co thickness \cite{Bandiera_2012, Carcia_1988, Zeper_1989}. Furthermore, a recent work by Denker \textit{et al.}\ on Ta/CoFeB/MgO films with variable CoFeB thickness yields comparable results, namely a transition from OOP to IP anisotropy toward higher thicknesses as indicated by similar hysteresis loops \cite{Denker_2020}.

\subsection{Temperature Dependence Of Magnetic Properties}
\begin{figure}
\centering
\includegraphics[width=7.0 cm]{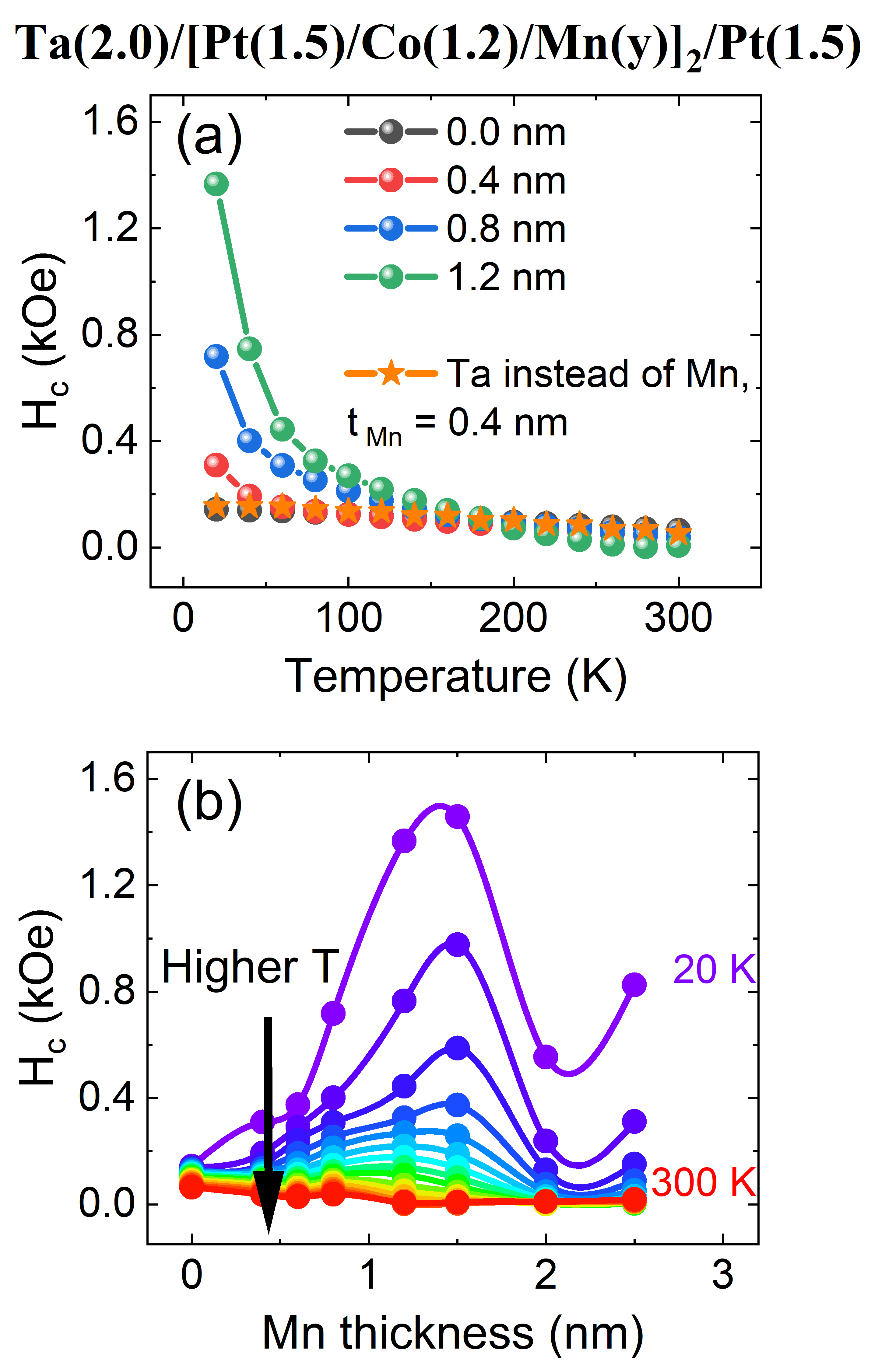}
\caption{(a) Temperature-dependent coercive field $H_{\mathrm{c}}$ determined for selected [Pt/Co/Mn]$_2$ samples with varying Mn thickness and a [Pt/Co/Ta]$_2$ reference sample between $20$ and $300$\,K. Toward higher Mn layer thicknesses, $H_{\mathrm{c}}$ increases drastically. (b) $H_{\mathrm{c}}$ plotted against the Mn thickness for different temperatures. A maximum is observed for samples with Mn thickness of $1.5$\,nm. Solid lines are guide to the eye.} 
\label{HCTEP}%
\end{figure}

The observed decreasing saturation magnetization for samples with higher Mn layer thickness served as a motivation for temperature-dependent measurements that could provide further insights about the nature of a possible antiferromagnetic coupling within the multilayer samples. In Fig.\ \ref{HCTEP}(a), the coercive field $H_{\mathrm{c}}$ for several Ta(2)/[Pt(1.5)/Co(1.2)/Mn($y$)]$_2$/Pt(1.5) samples with varying Mn thickness $y$ is plotted against temperature between $20$\,K and $300$\,K. Furthermore, a [Pt/Co/Ta]$_2$ reference sample with $0.4$\, nm thin layers of Ta instead of Mn has also been investigated. It can be seen that the [Pt/Co]$_2$ and the [Pt/Co/Ta]$_2$ multilayers do not exhibit any significant temperature dependence in $H_{\mathrm{c}}$. However, a pronounced increase toward low temperatures can be observed for samples containing Mn. In detail, this increase becomes more significant for the case of thicker Mn layers. As shown in Fig.\ \ref{HCTEP} (b), there is a maximum for Mn thicknesses between $1.0$\,nm and $1.5$\,nm, whereas the increase of $H_{\mathrm{c}}$ is less pronounced for thickness values outside of this range. 

The strong enhancement of the coercive field $H_{\mathrm{c}}$ for samples with intermediate Mn layer thicknesses upon cooling is in good agreement with the hypothesis of an emergent interfacial Co/Mn antiferromagnetic coupling. Our experimental data may also be explained by slight intermixing of Co and Mn and the emergence of a spin-glass-like phase of Co-Mn, in which a higher field (i.e., more energy) is required to rotate the frozen spins into the magnetic field direction. Such a spin-glass-like phase might even be accompanied by antiferromagnetic coupling, which is expected for Co-Mn alloys depending on the exact compositon \cite{Menshikov1985}. A pronounced increase of the coercive field toward low temperatures has been reported in a wide range of other materials due to various reasons. Selected works include amorphous Fe-Ni-B metallic glasses \cite{Thomas_2012} or rare-earth doped permalloy films \cite{Luo_2015}. While the first study by Thomas \textit{et al.}\ also discusses the existence of a spin-glass like phase, the increase in coercivity upon cooling for the latter materials system is explained by the film structure changing from polycrystalline to amorphous with increasing concentration of dopants. Such a structural transition from polycrystalline to more amorphous films is likewise conceivable for the Co-Mn interfacial region in the studied Pt/Co/Mn samples with varying Mn layer thickness.

Following the discussion on global magnetic properties of Pt/Co/Mn-based multilayers, we will now analyze the spatially-resolved magnetic states in these systems.

\subsection{Magneto-Optical Imaging}
\begin{figure*}
\centering
\includegraphics[width=17 cm]{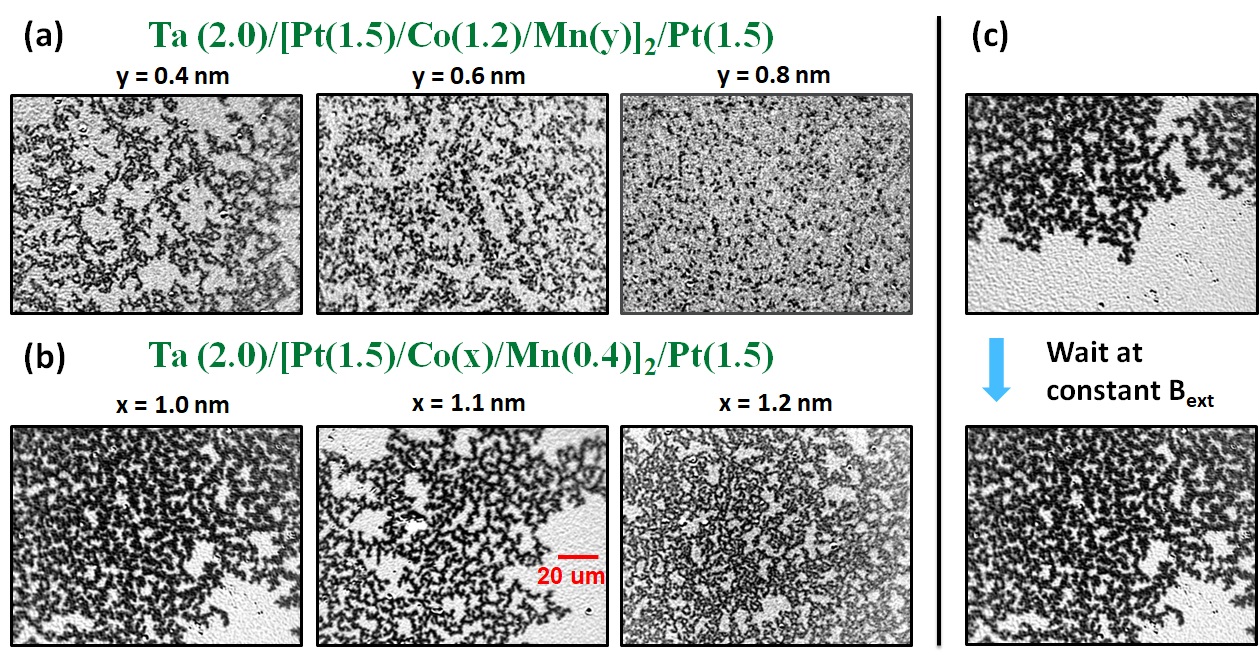}
\caption{MOKE images of [Pt/Co/Mn]$_2$ multilayers for selected samples with varying (a) Mn layer and (b) Co layer thickness. Scale bar (red color) corresponds to a length of $20\, \mathrm{\mu m}$. A sufficiently thick Mn layer leads to the presence of bubble domains or potentially skyrmions. (c) Example for domain structure changes within a time interval of $T=10$\,s at a constant external field $B_{\mathrm{ext}}\approx 20$\,Oe for a Ta(2)/[Pt(1.5)/Co(1)/Mn(0.4)]$_2$/Pt(1.5) specimen.} 
\label{MOKEIMAGE}%
\end{figure*}

Magneto-optical imaging of domains has been realized on a home-built Kerr microscope with support of advanced image processing algorithms which include the difference image technique as well as contrast stretching \cite{Schaefer_2007}. Due to the decreasing domain size for higher numbers $n$ of Pt/Co/Mn repetitions, we focus on a discussion of measurements for $n=2$. For samples with higher $n$, we did not observe any clear signatures of magnetic domains. It should also be emphasized that for samples with Mn layer thicknesses of $y\geq1.0$\,nm, the Kerr signal becomes increasingly weaker and thus magneto-optical imaging with adequate resolution is not possible with our current setup. Figure \ref{MOKEIMAGE}(a) shows representative images of samples with varying Mn thickness $y$ at zero external magnetic field after saturation with an out-of-plane field. We observe a dendritic domain structure at the lowest Mn thickness $y=0.4\,$nm. The average size of magnetic domains is clearly lower for the case of $y=0.6\,$nm. For an even higher Mn thickness of $y=0.8\,$nm, we mostly observe bubble-like domains, which could potentially be magnetic skyrmions that are formed as a consequence of the DMI introduced by the Mn layer. However, despite the expected presence of interfacial DMI due to the difference in Co and Mn work functions \cite{Ajejas2021}, the available experimental data is solely a hint and not an unambiguous proof for the bubble-like objects being skyrmions. A more detailed investigation of the static and dynamic properties of these bubble-like domains is beyond the scope of the present work.

In Fig.\ \ref{MOKEIMAGE} (b), three further representative images are shown for samples with varying Co thickness $x$ at zero external field after saturation. Here, the size of the domains likewise changes systematically as a function of the Co layer thickness, which is in concordance with MOKE studies of CoFeB films with varying thickness \cite{Denker_2020}.
Finally, Fig.\ \ref{MOKEIMAGE} (c) displays the interesting observation that the domain state changes as a function of time even at a constant magnetic field. This is attributed to the pinning and de-pinning of domain walls at defects which represent local minima in the complex energy landscape. Both MOKE images of the sample Ta(2)/[Pt(1.5)/Co(1)/Mn(0.4)]$_2$/Pt(1.5) are taken at an external magnetic field of approximately $20\,$Oe applied perpendicular to the film plane. The top image was taken right after the field value has been reached, whereas the bottom image was acquired after a waiting time of $10\,$s at a constant field value.
In the following, we will show that measurements of the total magnetic moment using SQUID magnetometry can provide further insights into this interesting behavior.

\begin{figure}
\centering
\includegraphics[width=8.5 cm]{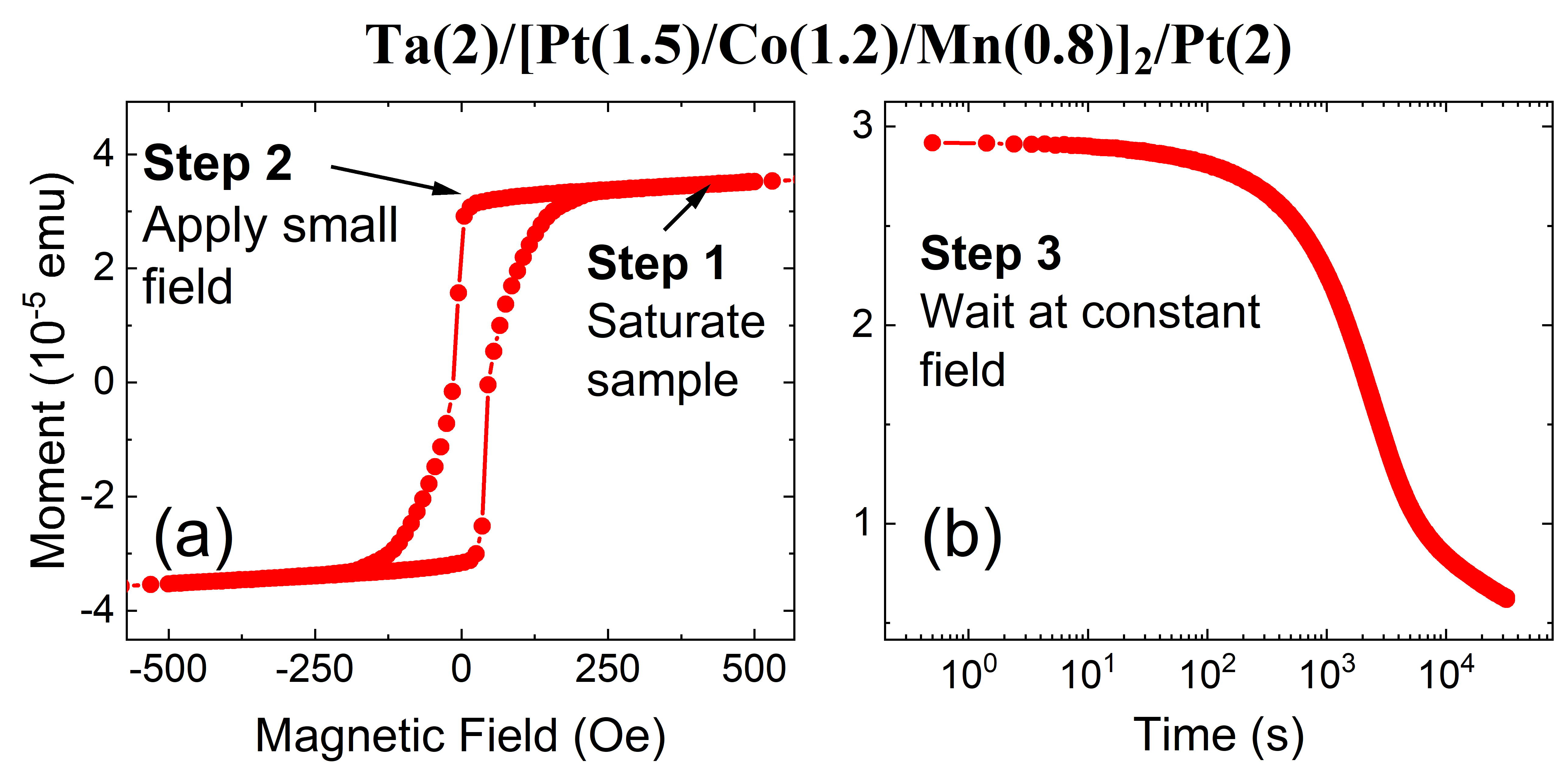}
\caption{Slow domain dynamics in a [Pt/Co/Mn]$_2$ sample at room temperature. (a) Magnetic hysteresis measured by SQUID/VSM. To observe the slow dynamics of domains, the sample is saturated with an out-of-plane field, which is then swept to a small value around zero. (b) In the last step, the time-dependent magnetization at this field is recorded for several hours. In a logarithmic representation, it follows an S-shape behavior.} 
\label{SLOWDYN}%
\end{figure}

The procedure for this type of experiment is summarized in Fig\ \ref{SLOWDYN}. Panel (a) displays the hysteresis of the sample. After saturation at a sufficiently large out-of-plane magnetic field, the field is swept to a small value around zero. Then, as depicted in Fig\ \ref{SLOWDYN}(b), the total magnetic moment is recorded for approximately nine hours at this constant small external field. When plotting the data on a logarithmic time axis, we observe an S-shaped curve. This has been found to be a universal behavior occuring in [Pt/Co/Mn]$_2$ samples with varying Mn thickness (not shown). Therefore, we attribute this slow domain dynamics to the characteristic structural properties of our fabricated multilayer stacks. We speculate that changes in the deposition conditions (e.g., varying the power and Ar partial pressure) may affect this behavior in a significant way. As the two selected MOKE images in Fig.\ \ref{MOKEIMAGE}(c) suggest, the domains exhibit a dendritic-like growth behavior, which is very similar to the observations in Ref.\ \cite{Woodward_2003} for the case of [Pt(1.86)/Co(0.36)]$_9$ multilayers. The resulting curve $M(t)$ for the Pt/Co multilayers studied by Woodward \textit{et al.}\ displays a similar behavior.   

\section{Summary and Conclusion}
In this work, we have experimentally characterized a novel Pt/Co-based multilayer system, namely [Pt/Co/Mn]$_n$. Structural investigations via XRR and STEM provide clear evidence for the high quality of the sputtered layers and also demonstrate the accuracy as well as reliability of the deposition process. Magnetic characterization by means of polar Kerr microscopy and SQUID/VSM magnetometry has revealed a systematic and complex behavior of this materials system. While the magnetic anisotropy can be tuned from out-of-plane to in-plane by increasing the Co layer thickness, a sufficiently thick Mn layer leads to the emergence of bubble-like magnetic domains -- potentially even skyrmions -- as well as to a reduction in the total magnetic moment.
The latter is ascribed to the emergence of interfacial Co/Mn antiferromagnetic coupling. Furthermore, a Co-Mn alloy possibly forms in a narrow interfacial region and may contain antiferromagnetic regions, which in the corresponding samples contribute to a strong increase of the coercive field at low temperatures. This behavior is most pronounced for Mn thicknesses around $1.5$\,nm and becomes less relevant for both higher and lower thickness values. Finally, we also observe a universal, time-dependent dendritic-like growth of magnetic domains in the Pt/Co/Mn materials system. 

We suggest the investigation of this materials system by means of Lorentz transmission electron microscopy or other comparable advanced imaging methods in order to reveal the character of the bubble-like domains unequivocally. In addition, such advanced imaging methods will allow for a more detailed characterization of multilayers with a high number of Pt/Co/Mn repetitions, where the average domain size is significantly smaller than for $n=2$ and thereby cannot be studied with our Kerr microscope. Alternatively, studying the current-driven dynamics of the observed bubble-like objects will also help to understand whether they are skyrmions or not.

In conclusion, it has been shown that tuning and understanding the magnetic properties of [Pt/Co/Mn]$_n$ requires careful engineering of and control over the structural properties of the individual metallic layers. While our STEM measurements do not indicate considerable intermixing of Co and Mn, it will be interesting to explore how even weak Co-Mn intermixing could be controlled more precisely, for instance by varying deposition parameters systematically and by inserting ultrathin Cu diffusion barriers as has been done for the case of [Co/Pt]$_n$ multilayers \cite{Bandiera_2012, Bandiera_2013} or by exploiting atomically thin two-dimensional materials such as graphene \cite{Morrow_2015, Hong_2014}.

\section*{Acknowledgements}
The authors would like to thank Fernando Ajejas for fruitful discussions on magnetic multilayers as well as Daniel Vaz for instructive comments on intermixing effects and diffusion barriers. Support during the sputtering process provided by Jonathan Gibbons and Saima Siddiqui is also gratefully acknowledged. M.\ L.\ acknowledges the financial support by the German Science Foundation (Deutsche Forschungsgemeinschaft, DFG) through the research fellowship LO 2584/1-1. The sample fabrication, structural and magnetic characterization, as well as data analysis and manuscript preparation was supported by the NSF through the University of Illinois at Urbana-Champaign Materials Research Science and Engineering Center DMR-1720633 and was carried out in part in the Materials Research Laboratory Central Research Facilities, University of Illinois. The transmission electron microscopy measurements were supported by the U.\ S.\ Department of Energy, Office of Science, Materials Science and Engineering Division under Contract No.\ DE-SC0022060.

\section*{Appendix A: Comparison of Different Magnetometry Techniques}
In order to verify the validity of the SQUID-magnetometry experiments, we compare the acquired hysteresis loops to MOKE measurements. In the latter experiment, we employ a phase sensitive detection method by means of a lock-in amplifier in order to optimize the signal-to-noise ratio \cite{Polisetty_2008}. In detail, the laser beam (wavelength $\lambda=633$\,nm) reflected from the multilayer sample is periodically modulated between left and right circularly polarized light by a photoelastic modulator with a frequency of $50$\,kHz. This signal is utilized as a reference signal for the lock-in amplifier, while the signal from a photosensitive diode corresponds to the input signal. 

\begin{figure}
\centering
\includegraphics[width=7 cm]{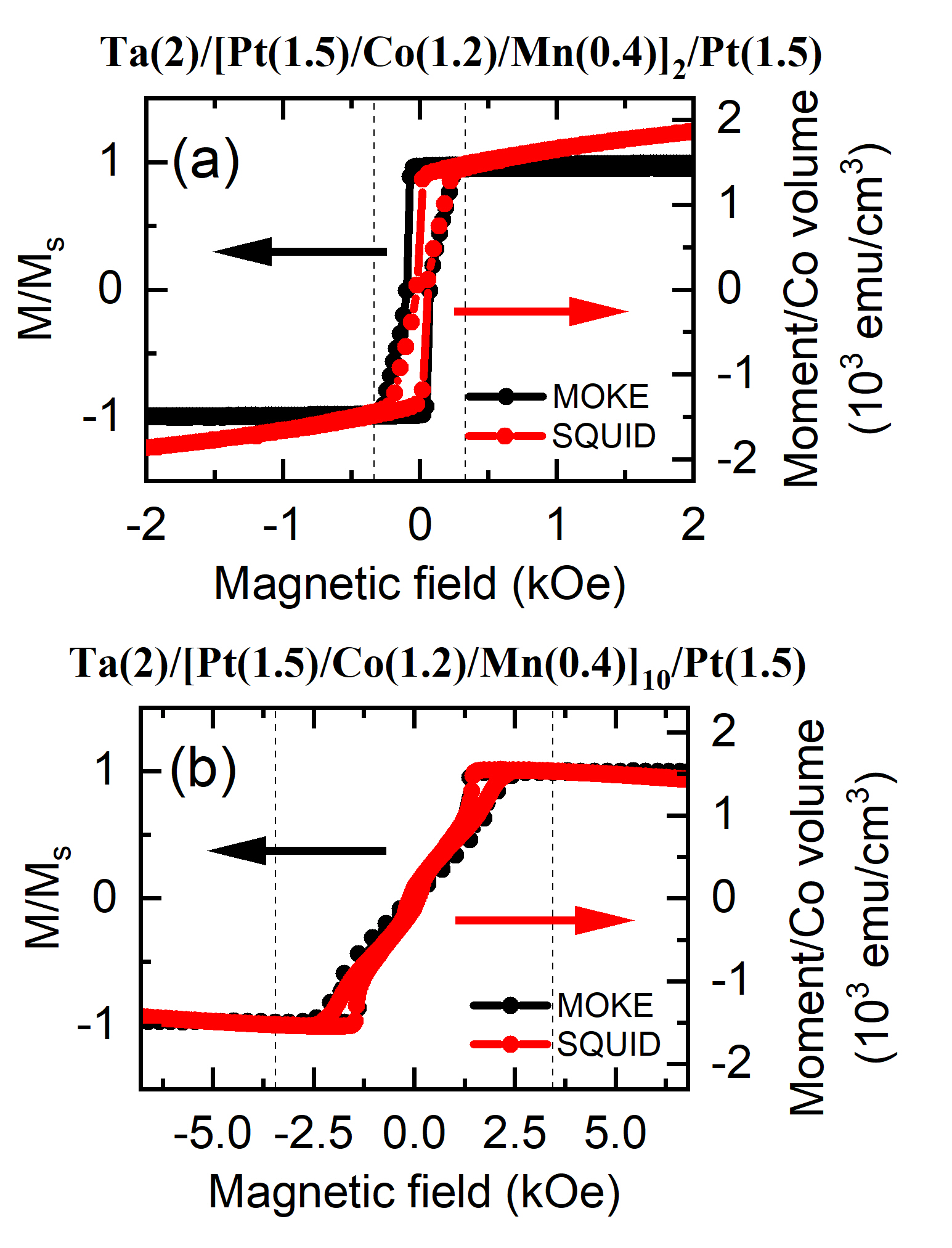}
\caption{Comparison of hysteresis loops for samples with (a) $n=2$ and (b) $n=10$ repetitions of Pt/Co/Mn recorded by two different experimental techniques: MOKE (black) and SQUID (red) magnetometry measurements. While the MOKE measurements do not exhibit any spurious background signal, the advantage of the SQUID measurement lies in the possibility of determining the magnetic moment of the entire multilayer stack.} 
\label{MOKESQUID}%
\end{figure}

Results for two representative [Pt/Co/Mn]$_n$ multilayers with $n=2$ and $n=10$ repeats are depicted in Fig.\ \ref{MOKESQUID}(a) and (b), respectively. Clearly, these diagrams demonstrate that both experimental techniques yield overlapping hysteresis curves for the most part. However, the MOKE measurements possess the advantage that no spurious background signal can be observed, whereas the curves recorded by SQUID magnetometry exhibit a non-zero slope even for external fields at which the sample is saturated. On the other hand, the SQUID measurements allow for determining the magnetic moment of the entire multilayer stack (right y-axis indicates the moment divided by the magnetic Co volume), while the proportionality factor between the Kerr rotation measured in the MOKE measurements and the total moment remains unknown. Therefore, we have only plotted the normalized magnetic moment $M/M_{\mathrm{s}}$ for this case (left y-axis). The SQUID experiments yield a magnetization density of around $1400\,$emu/cm$^3$ for the two depicted samples, which is in excellent agreement with the literature value for fcc cobalt \cite{Beaujour_2006}. Dashed vertical lines in the diagrams indicate the magnetic field at which the respective sample is saturated.

\bibliography{Skyrmion_literature}

\begin{thebibliography}{53}%
\makeatletter
\providecommand \@ifxundefined [1]{%
 \@ifx{#1\undefined}
}%
\providecommand \@ifnum [1]{%
 \ifnum #1\expandafter \@firstoftwo
 \else \expandafter \@secondoftwo
 \fi
}%
\providecommand \@ifx [1]{%
 \ifx #1\expandafter \@firstoftwo
 \else \expandafter \@secondoftwo
 \fi
}%
\providecommand \natexlab [1]{#1}%
\providecommand \enquote  [1]{``#1''}%
\providecommand \bibnamefont  [1]{#1}%
\providecommand \bibfnamefont [1]{#1}%
\providecommand \citenamefont [1]{#1}%
\providecommand \href@noop [0]{\@secondoftwo}%
\providecommand \href [0]{\begingroup \@sanitize@url \@href}%
\providecommand \@href[1]{\@@startlink{#1}\@@href}%
\providecommand \@@href[1]{\endgroup#1\@@endlink}%
\providecommand \@sanitize@url [0]{\catcode `\\12\catcode `\$12\catcode
  `\&12\catcode `\#12\catcode `\^12\catcode `\_12\catcode `\%12\relax}%
\providecommand \@@startlink[1]{}%
\providecommand \@@endlink[0]{}%
\providecommand \url  [0]{\begingroup\@sanitize@url \@url }%
\providecommand \@url [1]{\endgroup\@href {#1}{\urlprefix }}%
\providecommand \urlprefix  [0]{URL }%
\providecommand \Eprint [0]{\href }%
\providecommand \doibase [0]{http://dx.doi.org/}%
\providecommand \selectlanguage [0]{\@gobble}%
\providecommand \bibinfo  [0]{\@secondoftwo}%
\providecommand \bibfield  [0]{\@secondoftwo}%
\providecommand \translation [1]{[#1]}%
\providecommand \BibitemOpen [0]{}%
\providecommand \bibitemStop [0]{}%
\providecommand \bibitemNoStop [0]{.\EOS\space}%
\providecommand \EOS [0]{\spacefactor3000\relax}%
\providecommand \BibitemShut  [1]{\csname bibitem#1\endcsname}%
\let\auto@bib@innerbib\@empty
\bibitem [{\citenamefont {Schuller}\ \emph {et~al.}(1999)\citenamefont
  {Schuller}, \citenamefont {Kim},\ and\ \citenamefont
  {Leighton}}]{Schuller_1999}%
  \BibitemOpen
  \bibfield  {author} {\bibinfo {author} {\bibfnamefont {I.~K.}\ \bibnamefont
  {Schuller}}, \bibinfo {author} {\bibfnamefont {S.}~\bibnamefont {Kim}}, \
  and\ \bibinfo {author} {\bibfnamefont {C.}~\bibnamefont {Leighton}},\ }\href
  {\doibase 10.1016/S0304-8853(99)00336-4} {\bibfield  {journal} {\bibinfo
  {journal} {Journal of Magnetism and Magnetic Materials}\ }\textbf {\bibinfo
  {volume} {200}},\ \bibinfo {pages} {571} (\bibinfo {year}
  {1999})}\BibitemShut {NoStop}%
\bibitem [{\citenamefont {Bennett}\ and\ \citenamefont
  {Watson}(1994)}]{Bennett_1994}%
  \BibitemOpen
  \bibfield  {author} {\bibinfo {author} {\bibfnamefont {L.~H.}\ \bibnamefont
  {Bennett}}\ and\ \bibinfo {author} {\bibfnamefont {R.~E.}\ \bibnamefont
  {Watson}},\ }\href {\doibase 10.1142/2359} {\bibfield  {journal} {\bibinfo
  {journal} {World Scientific}\ } (\bibinfo {year} {1994}),\
  10.1142/2359}\BibitemShut {NoStop}%
\bibitem [{\citenamefont {Camley}\ and\ \citenamefont
  {Stamps}(1993)}]{Camley_1993}%
  \BibitemOpen
  \bibfield  {author} {\bibinfo {author} {\bibfnamefont {R.~E.}\ \bibnamefont
  {Camley}}\ and\ \bibinfo {author} {\bibfnamefont {R.~L.}\ \bibnamefont
  {Stamps}},\ }\href {\doibase 10.1088/0953-8984/5/23/003} {\bibfield
  {journal} {\bibinfo  {journal} {Journal of Physics: Condensed Matter}\
  }\textbf {\bibinfo {volume} {5}},\ \bibinfo {pages} {3727} (\bibinfo {year}
  {1993})}\BibitemShut {NoStop}%
\bibitem [{\citenamefont {Vedmedenko}\ \emph {et~al.}(2020)\citenamefont
  {Vedmedenko}, \citenamefont {Kawakami}, \citenamefont {Sheka}, \citenamefont
  {Gambardella}, \citenamefont {Kirilyuk}, \citenamefont {Hirohata},
  \citenamefont {Binek}, \citenamefont {Chubykalo-Fesenko}, \citenamefont
  {Sanvito}, \citenamefont {Kirby}, \citenamefont {Grollier}, \citenamefont
  {Everschor-Sitte}, \citenamefont {Kampfrath}, \citenamefont {You},\ and\
  \citenamefont {Berger}}]{Vedmedenko_2020}%
  \BibitemOpen
  \bibfield  {author} {\bibinfo {author} {\bibfnamefont {E.~Y.}\ \bibnamefont
  {Vedmedenko}}, \bibinfo {author} {\bibfnamefont {R.~K.}\ \bibnamefont
  {Kawakami}}, \bibinfo {author} {\bibfnamefont {D.~D.}\ \bibnamefont {Sheka}},
  \bibinfo {author} {\bibfnamefont {P.}~\bibnamefont {Gambardella}}, \bibinfo
  {author} {\bibfnamefont {A.}~\bibnamefont {Kirilyuk}}, \bibinfo {author}
  {\bibfnamefont {A.}~\bibnamefont {Hirohata}}, \bibinfo {author}
  {\bibfnamefont {C.}~\bibnamefont {Binek}}, \bibinfo {author} {\bibfnamefont
  {O.}~\bibnamefont {Chubykalo-Fesenko}}, \bibinfo {author} {\bibfnamefont
  {S.}~\bibnamefont {Sanvito}}, \bibinfo {author} {\bibfnamefont {B.~J.}\
  \bibnamefont {Kirby}}, \bibinfo {author} {\bibfnamefont {J.}~\bibnamefont
  {Grollier}}, \bibinfo {author} {\bibfnamefont {K.}~\bibnamefont
  {Everschor-Sitte}}, \bibinfo {author} {\bibfnamefont {T.}~\bibnamefont
  {Kampfrath}}, \bibinfo {author} {\bibfnamefont {C.-Y.}\ \bibnamefont {You}},
  \ and\ \bibinfo {author} {\bibfnamefont {A.}~\bibnamefont {Berger}},\ }\href
  {\doibase 10.1088/1361-6463/ab9d98} {\bibfield  {journal} {\bibinfo
  {journal} {Journal of Physics D: Applied Physics}\ }\textbf {\bibinfo
  {volume} {53}},\ \bibinfo {pages} {453001} (\bibinfo {year}
  {2020})}\BibitemShut {NoStop}%
\bibitem [{\citenamefont {Back}\ \emph {et~al.}(2020)\citenamefont {Back},
  \citenamefont {Cros}, \citenamefont {Ebert}, \citenamefont {Everschor-Sitte},
  \citenamefont {Fert}, \citenamefont {Garst}, \citenamefont {Ma},
  \citenamefont {Mankovsky}, \citenamefont {Monchesky}, \citenamefont
  {Mostovoy}, \citenamefont {Nagaosa}, \citenamefont {Parkin}, \citenamefont
  {Pfleiderer}, \citenamefont {Reyren}, \citenamefont {Rosch}, \citenamefont
  {Taguchi}, \citenamefont {Tokura}, \citenamefont {von Bergmann},\ and\
  \citenamefont {Zang}}]{Back2020}%
  \BibitemOpen
  \bibfield  {author} {\bibinfo {author} {\bibfnamefont {C.}~\bibnamefont
  {Back}}, \bibinfo {author} {\bibfnamefont {V.}~\bibnamefont {Cros}}, \bibinfo
  {author} {\bibfnamefont {H.}~\bibnamefont {Ebert}}, \bibinfo {author}
  {\bibfnamefont {K.}~\bibnamefont {Everschor-Sitte}}, \bibinfo {author}
  {\bibfnamefont {A.}~\bibnamefont {Fert}}, \bibinfo {author} {\bibfnamefont
  {M.}~\bibnamefont {Garst}}, \bibinfo {author} {\bibfnamefont
  {T.}~\bibnamefont {Ma}}, \bibinfo {author} {\bibfnamefont {S.}~\bibnamefont
  {Mankovsky}}, \bibinfo {author} {\bibfnamefont {T.~L.}\ \bibnamefont
  {Monchesky}}, \bibinfo {author} {\bibfnamefont {M.}~\bibnamefont {Mostovoy}},
  \bibinfo {author} {\bibfnamefont {N.}~\bibnamefont {Nagaosa}}, \bibinfo
  {author} {\bibfnamefont {S.~S.~P.}\ \bibnamefont {Parkin}}, \bibinfo {author}
  {\bibfnamefont {C.}~\bibnamefont {Pfleiderer}}, \bibinfo {author}
  {\bibfnamefont {N.}~\bibnamefont {Reyren}}, \bibinfo {author} {\bibfnamefont
  {A.}~\bibnamefont {Rosch}}, \bibinfo {author} {\bibfnamefont
  {Y.}~\bibnamefont {Taguchi}}, \bibinfo {author} {\bibfnamefont
  {Y.}~\bibnamefont {Tokura}}, \bibinfo {author} {\bibfnamefont
  {K.}~\bibnamefont {von Bergmann}}, \ and\ \bibinfo {author} {\bibfnamefont
  {J.}~\bibnamefont {Zang}},\ }\href {\doibase 10.1088/1361-6463} {\bibfield
  {journal} {\bibinfo  {journal} {Journal of Physics D: Applied Physics}\
  }\textbf {\bibinfo {volume} {53}},\ \bibinfo {pages} {363001} (\bibinfo
  {year} {2020})}\BibitemShut {NoStop}%
\bibitem [{\citenamefont {Hellman}\ \emph {et~al.}(2017)\citenamefont
  {Hellman}, \citenamefont {Hoffmann}, \citenamefont {Tserkovnyak},
  \citenamefont {Beach}, \citenamefont {Fullerton}, \citenamefont {Leighton},
  \citenamefont {MacDonald}, \citenamefont {Ralph}, \citenamefont {Arena},
  \citenamefont {Dürr}, \citenamefont {Fischer}, \citenamefont {Grollier},
  \citenamefont {Heremans}, \citenamefont {Jungwirth}, \citenamefont {Kimel},
  \citenamefont {Koopmans}, \citenamefont {Krivorotov}, \citenamefont {May},
  \citenamefont {Petford-Long}, \citenamefont {Rondinelli}, \citenamefont
  {Samarth}, \citenamefont {Schuller}, \citenamefont {Slavin}, \citenamefont
  {Stiles}, \citenamefont {Tchernyshyov}, \citenamefont {Thiaville},\ and\
  \citenamefont {Zink}}]{Hellman_2017}%
  \BibitemOpen
  \bibfield  {author} {\bibinfo {author} {\bibfnamefont {F.}~\bibnamefont
  {Hellman}}, \bibinfo {author} {\bibfnamefont {A.}~\bibnamefont {Hoffmann}},
  \bibinfo {author} {\bibfnamefont {Y.}~\bibnamefont {Tserkovnyak}}, \bibinfo
  {author} {\bibfnamefont {G.~S.}\ \bibnamefont {Beach}}, \bibinfo {author}
  {\bibfnamefont {E.~E.}\ \bibnamefont {Fullerton}}, \bibinfo {author}
  {\bibfnamefont {C.}~\bibnamefont {Leighton}}, \bibinfo {author}
  {\bibfnamefont {A.~H.}\ \bibnamefont {MacDonald}}, \bibinfo {author}
  {\bibfnamefont {D.~C.}\ \bibnamefont {Ralph}}, \bibinfo {author}
  {\bibfnamefont {D.~A.}\ \bibnamefont {Arena}}, \bibinfo {author}
  {\bibfnamefont {H.~A.}\ \bibnamefont {Dürr}}, \bibinfo {author}
  {\bibfnamefont {P.}~\bibnamefont {Fischer}}, \bibinfo {author} {\bibfnamefont
  {J.}~\bibnamefont {Grollier}}, \bibinfo {author} {\bibfnamefont {J.~P.}\
  \bibnamefont {Heremans}}, \bibinfo {author} {\bibfnamefont {T.}~\bibnamefont
  {Jungwirth}}, \bibinfo {author} {\bibfnamefont {A.~V.}\ \bibnamefont
  {Kimel}}, \bibinfo {author} {\bibfnamefont {B.}~\bibnamefont {Koopmans}},
  \bibinfo {author} {\bibfnamefont {I.~N.}\ \bibnamefont {Krivorotov}},
  \bibinfo {author} {\bibfnamefont {S.~J.}\ \bibnamefont {May}}, \bibinfo
  {author} {\bibfnamefont {A.~K.}\ \bibnamefont {Petford-Long}}, \bibinfo
  {author} {\bibfnamefont {J.~M.}\ \bibnamefont {Rondinelli}}, \bibinfo
  {author} {\bibfnamefont {N.}~\bibnamefont {Samarth}}, \bibinfo {author}
  {\bibfnamefont {I.~K.}\ \bibnamefont {Schuller}}, \bibinfo {author}
  {\bibfnamefont {A.~N.}\ \bibnamefont {Slavin}}, \bibinfo {author}
  {\bibfnamefont {M.~D.}\ \bibnamefont {Stiles}}, \bibinfo {author}
  {\bibfnamefont {O.}~\bibnamefont {Tchernyshyov}}, \bibinfo {author}
  {\bibfnamefont {A.}~\bibnamefont {Thiaville}}, \ and\ \bibinfo {author}
  {\bibfnamefont {B.~L.}\ \bibnamefont {Zink}},\ }\href {\doibase
  10.1103/RevModPhys.89.025006} {\bibfield  {journal} {\bibinfo  {journal}
  {Reviews of Modern Physics}\ }\textbf {\bibinfo {volume} {89}},\ \bibinfo
  {pages} {025006} (\bibinfo {year} {2017})}\BibitemShut {NoStop}%
\bibitem [{\citenamefont {Dzyaloshinskii}(1958)}]{Dzyaloshinskii1958}%
  \BibitemOpen
  \bibfield  {author} {\bibinfo {author} {\bibfnamefont {I.}~\bibnamefont
  {Dzyaloshinskii}},\ }\href {\doibase 10.1016/0022-3697(58)90076-3} {\bibfield
   {journal} {\bibinfo  {journal} {Journal of Physics and Chemistry of Solids}\
  }\textbf {\bibinfo {volume} {4}},\ \bibinfo {pages} {241} (\bibinfo {year}
  {1958})}\BibitemShut {NoStop}%
\bibitem [{\citenamefont {Moriya}(1960)}]{Moriya1960}%
  \BibitemOpen
  \bibfield  {author} {\bibinfo {author} {\bibfnamefont {T.}~\bibnamefont
  {Moriya}},\ }\href {\doibase 10.1103/PhysRev.120.91} {\bibfield  {journal}
  {\bibinfo  {journal} {Physical Review}\ }\textbf {\bibinfo {volume} {120}},\
  \bibinfo {pages} {91} (\bibinfo {year} {1960})}\BibitemShut {NoStop}%
\bibitem [{\citenamefont {Jiang}\ \emph {et~al.}(2017)\citenamefont {Jiang},
  \citenamefont {Chen}, \citenamefont {Liu}, \citenamefont {Zang},
  \citenamefont {te~Velthuis},\ and\ \citenamefont {Hoffmann}}]{Jiang2017}%
  \BibitemOpen
  \bibfield  {author} {\bibinfo {author} {\bibfnamefont {W.}~\bibnamefont
  {Jiang}}, \bibinfo {author} {\bibfnamefont {G.}~\bibnamefont {Chen}},
  \bibinfo {author} {\bibfnamefont {K.}~\bibnamefont {Liu}}, \bibinfo {author}
  {\bibfnamefont {J.}~\bibnamefont {Zang}}, \bibinfo {author} {\bibfnamefont
  {S.~G.}\ \bibnamefont {te~Velthuis}}, \ and\ \bibinfo {author} {\bibfnamefont
  {A.}~\bibnamefont {Hoffmann}},\ }\href {\doibase
  10.1016/j.physrep.2017.08.001} {\bibfield  {journal} {\bibinfo  {journal}
  {Physics Reports}\ }\textbf {\bibinfo {volume} {704}},\ \bibinfo {pages} {1}
  (\bibinfo {year} {2017})}\BibitemShut {NoStop}%
\bibitem [{\citenamefont {Woo}\ \emph {et~al.}(2016)\citenamefont {Woo},
  \citenamefont {Litzius}, \citenamefont {Krüger}, \citenamefont {Im},
  \citenamefont {Caretta}, \citenamefont {Richter}, \citenamefont {Mann},
  \citenamefont {Krone}, \citenamefont {Reeve}, \citenamefont {Weigand},
  \citenamefont {Agrawal}, \citenamefont {Lemesh}, \citenamefont {Mawass},
  \citenamefont {Fischer}, \citenamefont {Kläui},\ and\ \citenamefont
  {Beach}}]{Woo2016}%
  \BibitemOpen
  \bibfield  {author} {\bibinfo {author} {\bibfnamefont {S.}~\bibnamefont
  {Woo}}, \bibinfo {author} {\bibfnamefont {K.}~\bibnamefont {Litzius}},
  \bibinfo {author} {\bibfnamefont {B.}~\bibnamefont {Krüger}}, \bibinfo
  {author} {\bibfnamefont {M.-Y.}\ \bibnamefont {Im}}, \bibinfo {author}
  {\bibfnamefont {L.}~\bibnamefont {Caretta}}, \bibinfo {author} {\bibfnamefont
  {K.}~\bibnamefont {Richter}}, \bibinfo {author} {\bibfnamefont
  {M.}~\bibnamefont {Mann}}, \bibinfo {author} {\bibfnamefont {A.}~\bibnamefont
  {Krone}}, \bibinfo {author} {\bibfnamefont {R.~M.}\ \bibnamefont {Reeve}},
  \bibinfo {author} {\bibfnamefont {M.}~\bibnamefont {Weigand}}, \bibinfo
  {author} {\bibfnamefont {P.}~\bibnamefont {Agrawal}}, \bibinfo {author}
  {\bibfnamefont {I.}~\bibnamefont {Lemesh}}, \bibinfo {author} {\bibfnamefont
  {M.-A.}\ \bibnamefont {Mawass}}, \bibinfo {author} {\bibfnamefont
  {P.}~\bibnamefont {Fischer}}, \bibinfo {author} {\bibfnamefont
  {M.}~\bibnamefont {Kläui}}, \ and\ \bibinfo {author} {\bibfnamefont
  {G.~S.~D.}\ \bibnamefont {Beach}},\ }\href {\doibase 10.1038/nmat4593}
  {\bibfield  {journal} {\bibinfo  {journal} {Nature Materials}\ }\textbf
  {\bibinfo {volume} {15}},\ \bibinfo {pages} {501} (\bibinfo {year}
  {2016})}\BibitemShut {NoStop}%
\bibitem [{\citenamefont {Woo}\ \emph {et~al.}(2018)\citenamefont {Woo},
  \citenamefont {Song}, \citenamefont {Zhang}, \citenamefont {Ezawa},
  \citenamefont {Zhou}, \citenamefont {Liu}, \citenamefont {Weigand},
  \citenamefont {Finizio}, \citenamefont {Raabe}, \citenamefont {Park},
  \citenamefont {Lee}, \citenamefont {Choi}, \citenamefont {Min}, \citenamefont
  {Koo},\ and\ \citenamefont {Chang}}]{Woo_2018}%
  \BibitemOpen
  \bibfield  {author} {\bibinfo {author} {\bibfnamefont {S.}~\bibnamefont
  {Woo}}, \bibinfo {author} {\bibfnamefont {K.~M.}\ \bibnamefont {Song}},
  \bibinfo {author} {\bibfnamefont {X.}~\bibnamefont {Zhang}}, \bibinfo
  {author} {\bibfnamefont {M.}~\bibnamefont {Ezawa}}, \bibinfo {author}
  {\bibfnamefont {Y.}~\bibnamefont {Zhou}}, \bibinfo {author} {\bibfnamefont
  {X.}~\bibnamefont {Liu}}, \bibinfo {author} {\bibfnamefont {M.}~\bibnamefont
  {Weigand}}, \bibinfo {author} {\bibfnamefont {S.}~\bibnamefont {Finizio}},
  \bibinfo {author} {\bibfnamefont {J.}~\bibnamefont {Raabe}}, \bibinfo
  {author} {\bibfnamefont {M.-C.}\ \bibnamefont {Park}}, \bibinfo {author}
  {\bibfnamefont {K.-Y.}\ \bibnamefont {Lee}}, \bibinfo {author} {\bibfnamefont
  {J.~W.}\ \bibnamefont {Choi}}, \bibinfo {author} {\bibfnamefont {B.-C.}\
  \bibnamefont {Min}}, \bibinfo {author} {\bibfnamefont {H.~C.}\ \bibnamefont
  {Koo}}, \ and\ \bibinfo {author} {\bibfnamefont {J.}~\bibnamefont {Chang}},\
  }\href {\doibase 10.1038/s41928-018-0070-8} {\bibfield  {journal} {\bibinfo
  {journal} {Nature Electronics}\ }\textbf {\bibinfo {volume} {1}},\ \bibinfo
  {pages} {288} (\bibinfo {year} {2018})}\BibitemShut {NoStop}%
\bibitem [{\citenamefont {Everschor-Sitte}\ \emph {et~al.}(2018)\citenamefont
  {Everschor-Sitte}, \citenamefont {Masell}, \citenamefont {Reeve},\ and\
  \citenamefont {Kläui}}]{Everschor-Sitte2018}%
  \BibitemOpen
  \bibfield  {author} {\bibinfo {author} {\bibfnamefont {K.}~\bibnamefont
  {Everschor-Sitte}}, \bibinfo {author} {\bibfnamefont {J.}~\bibnamefont
  {Masell}}, \bibinfo {author} {\bibfnamefont {R.~M.}\ \bibnamefont {Reeve}}, \
  and\ \bibinfo {author} {\bibfnamefont {M.}~\bibnamefont {Kläui}},\ }\href
  {\doibase 10.1063/1.5048972} {\bibfield  {journal} {\bibinfo  {journal}
  {Journal of Applied Physics}\ }\textbf {\bibinfo {volume} {124}},\ \bibinfo
  {pages} {240901} (\bibinfo {year} {2018})}\BibitemShut {NoStop}%
\bibitem [{\citenamefont {Kim}\ \emph {et~al.}(2017)\citenamefont {Kim},
  \citenamefont {Yang}, \citenamefont {Cho}, \citenamefont {Kim},\ and\
  \citenamefont {Kim}}]{Kim2017}%
  \BibitemOpen
  \bibfield  {author} {\bibinfo {author} {\bibfnamefont {J.}~\bibnamefont
  {Kim}}, \bibinfo {author} {\bibfnamefont {J.}~\bibnamefont {Yang}}, \bibinfo
  {author} {\bibfnamefont {Y.-J.}\ \bibnamefont {Cho}}, \bibinfo {author}
  {\bibfnamefont {B.}~\bibnamefont {Kim}}, \ and\ \bibinfo {author}
  {\bibfnamefont {S.-K.}\ \bibnamefont {Kim}},\ }\href {\doibase
  10.1038/srep45185} {\bibfield  {journal} {\bibinfo  {journal} {Scientific
  Reports}\ }\textbf {\bibinfo {volume} {7}},\ \bibinfo {pages} {45185}
  (\bibinfo {year} {2017})}\BibitemShut {NoStop}%
\bibitem [{\citenamefont {Kim}\ \emph {et~al.}(2018)\citenamefont {Kim},
  \citenamefont {Yang}, \citenamefont {Cho}, \citenamefont {Kim},\ and\
  \citenamefont {Kim}}]{Kim2018}%
  \BibitemOpen
  \bibfield  {author} {\bibinfo {author} {\bibfnamefont {J.}~\bibnamefont
  {Kim}}, \bibinfo {author} {\bibfnamefont {J.}~\bibnamefont {Yang}}, \bibinfo
  {author} {\bibfnamefont {Y.-J.}\ \bibnamefont {Cho}}, \bibinfo {author}
  {\bibfnamefont {B.}~\bibnamefont {Kim}}, \ and\ \bibinfo {author}
  {\bibfnamefont {S.-K.}\ \bibnamefont {Kim}},\ }\href {\doibase
  10.1063/1.5010948} {\bibfield  {journal} {\bibinfo  {journal} {Journal of
  Applied Physics}\ }\textbf {\bibinfo {volume} {123}},\ \bibinfo {pages}
  {053903} (\bibinfo {year} {2018})}\BibitemShut {NoStop}%
\bibitem [{\citenamefont {Lonsky}\ and\ \citenamefont
  {Hoffmann}(2020{\natexlab{a}})}]{Lonsky2020a}%
  \BibitemOpen
  \bibfield  {author} {\bibinfo {author} {\bibfnamefont {M.}~\bibnamefont
  {Lonsky}}\ and\ \bibinfo {author} {\bibfnamefont {A.}~\bibnamefont
  {Hoffmann}},\ }\href {\doibase 10.1063/5.0027042} {\bibfield  {journal}
  {\bibinfo  {journal} {{APL} Materials}\ }\textbf {\bibinfo {volume} {8}},\
  \bibinfo {pages} {100903} (\bibinfo {year} {2020}{\natexlab{a}})}\BibitemShut
  {NoStop}%
\bibitem [{\citenamefont {Lonsky}\ and\ \citenamefont
  {Hoffmann}(2020{\natexlab{b}})}]{Lonsky2020}%
  \BibitemOpen
  \bibfield  {author} {\bibinfo {author} {\bibfnamefont {M.}~\bibnamefont
  {Lonsky}}\ and\ \bibinfo {author} {\bibfnamefont {A.}~\bibnamefont
  {Hoffmann}},\ }\href {\doibase 10.1103/physrevb.102.104403} {\bibfield
  {journal} {\bibinfo  {journal} {Physical Review B}\ }\textbf {\bibinfo
  {volume} {102}},\ \bibinfo {pages} {104403} (\bibinfo {year}
  {2020}{\natexlab{b}})}\BibitemShut {NoStop}%
\bibitem [{\citenamefont {Chen}\ and\ \citenamefont {Ma}(2021)}]{Chen_2021}%
  \BibitemOpen
  \bibfield  {author} {\bibinfo {author} {\bibfnamefont {Z.}~\bibnamefont
  {Chen}}\ and\ \bibinfo {author} {\bibfnamefont {F.}~\bibnamefont {Ma}},\
  }\href {\doibase 10.1063/5.0061832} {\bibfield  {journal} {\bibinfo
  {journal} {Journal of Applied Physics}\ }\textbf {\bibinfo {volume} {130}},\
  \bibinfo {pages} {090901} (\bibinfo {year} {2021})}\BibitemShut {NoStop}%
\bibitem [{\citenamefont {Jia}\ \emph {et~al.}(2020)\citenamefont {Jia},
  \citenamefont {Zimmermann}, \citenamefont {Hoffmann}, \citenamefont
  {Sallermann}, \citenamefont {Bihlmayer},\ and\ \citenamefont
  {Blügel}}]{Jia2020}%
  \BibitemOpen
  \bibfield  {author} {\bibinfo {author} {\bibfnamefont {H.}~\bibnamefont
  {Jia}}, \bibinfo {author} {\bibfnamefont {B.}~\bibnamefont {Zimmermann}},
  \bibinfo {author} {\bibfnamefont {M.}~\bibnamefont {Hoffmann}}, \bibinfo
  {author} {\bibfnamefont {M.}~\bibnamefont {Sallermann}}, \bibinfo {author}
  {\bibfnamefont {G.}~\bibnamefont {Bihlmayer}}, \ and\ \bibinfo {author}
  {\bibfnamefont {S.}~\bibnamefont {Blügel}},\ }\href {\doibase
  10.1103/PhysRevMaterials.4.094407} {\bibfield  {journal} {\bibinfo  {journal}
  {Physical Review Materials}\ }\textbf {\bibinfo {volume} {4}},\ \bibinfo
  {pages} {094407} (\bibinfo {year} {2020})}\BibitemShut {NoStop}%
\bibitem [{\citenamefont {Ajejas}\ \emph {et~al.}(2021)\citenamefont {Ajejas},
  \citenamefont {Sassi}, \citenamefont {Legrand}, \citenamefont {Collin},
  \citenamefont {Thiaville}, \citenamefont {Garcia}, \citenamefont {Pizzini},
  \citenamefont {Reyren}, \citenamefont {Cros},\ and\ \citenamefont
  {Fert}}]{Ajejas2021}%
  \BibitemOpen
  \bibfield  {author} {\bibinfo {author} {\bibfnamefont {F.}~\bibnamefont
  {Ajejas}}, \bibinfo {author} {\bibfnamefont {Y.}~\bibnamefont {Sassi}},
  \bibinfo {author} {\bibfnamefont {W.}~\bibnamefont {Legrand}}, \bibinfo
  {author} {\bibfnamefont {S.}~\bibnamefont {Collin}}, \bibinfo {author}
  {\bibfnamefont {A.}~\bibnamefont {Thiaville}}, \bibinfo {author}
  {\bibfnamefont {J.~P.}\ \bibnamefont {Garcia}}, \bibinfo {author}
  {\bibfnamefont {S.}~\bibnamefont {Pizzini}}, \bibinfo {author} {\bibfnamefont
  {N.}~\bibnamefont {Reyren}}, \bibinfo {author} {\bibfnamefont
  {V.}~\bibnamefont {Cros}}, \ and\ \bibinfo {author} {\bibfnamefont
  {A.}~\bibnamefont {Fert}},\ }\href@noop {} {\  (\bibinfo {year} {2021})},\
  \Eprint {http://arxiv.org/abs/https://arxiv.org/abs/2109.00761}
  {arXiv:https://arxiv.org/abs/2109.00761 [cond-mat.mtrl-sci]} \BibitemShut
  {NoStop}%
\bibitem [{\citenamefont {Wells}\ \emph {et~al.}(2017)\citenamefont {Wells},
  \citenamefont {Shepley}, \citenamefont {Marrows},\ and\ \citenamefont
  {Moore}}]{Wells2017}%
  \BibitemOpen
  \bibfield  {author} {\bibinfo {author} {\bibfnamefont {A.~W.~J.}\
  \bibnamefont {Wells}}, \bibinfo {author} {\bibfnamefont {P.~M.}\ \bibnamefont
  {Shepley}}, \bibinfo {author} {\bibfnamefont {C.~H.}\ \bibnamefont
  {Marrows}}, \ and\ \bibinfo {author} {\bibfnamefont {T.~A.}\ \bibnamefont
  {Moore}},\ }\href {\doibase 10.1103/PhysRevB.95.054428} {\bibfield  {journal}
  {\bibinfo  {journal} {Physical Review B}\ }\textbf {\bibinfo {volume} {95}},\
  \bibinfo {pages} {054428} (\bibinfo {year} {2017})}\BibitemShut {NoStop}%
\bibitem [{\citenamefont {Brock}\ \emph {et~al.}(2020)\citenamefont {Brock},
  \citenamefont {Montoya}, \citenamefont {Im},\ and\ \citenamefont
  {Fullerton}}]{Brock2020}%
  \BibitemOpen
  \bibfield  {author} {\bibinfo {author} {\bibfnamefont {J.~A.}\ \bibnamefont
  {Brock}}, \bibinfo {author} {\bibfnamefont {S.~A.}\ \bibnamefont {Montoya}},
  \bibinfo {author} {\bibfnamefont {M.-Y.}\ \bibnamefont {Im}}, \ and\ \bibinfo
  {author} {\bibfnamefont {E.~E.}\ \bibnamefont {Fullerton}},\ }\href {\doibase
  10.1103/PhysRevMaterials.4.104409} {\bibfield  {journal} {\bibinfo  {journal}
  {Physical Review Materials}\ }\textbf {\bibinfo {volume} {4}},\ \bibinfo
  {pages} {104409} (\bibinfo {year} {2020})}\BibitemShut {NoStop}%
\bibitem [{\citenamefont {Jiang}\ \emph {et~al.}(2019)\citenamefont {Jiang},
  \citenamefont {Zhang}, \citenamefont {Wang}, \citenamefont {Phatak},
  \citenamefont {Wang}, \citenamefont {Zhang}, \citenamefont {Jungfleisch},
  \citenamefont {Pearson}, \citenamefont {Liu}, \citenamefont {Zang},
  \citenamefont {Cheng}, \citenamefont {Petford-Long}, \citenamefont
  {Hoffmann},\ and\ \citenamefont {te~Velthuis}}]{Jiang2019}%
  \BibitemOpen
  \bibfield  {author} {\bibinfo {author} {\bibfnamefont {W.}~\bibnamefont
  {Jiang}}, \bibinfo {author} {\bibfnamefont {S.}~\bibnamefont {Zhang}},
  \bibinfo {author} {\bibfnamefont {X.}~\bibnamefont {Wang}}, \bibinfo {author}
  {\bibfnamefont {C.}~\bibnamefont {Phatak}}, \bibinfo {author} {\bibfnamefont
  {Q.}~\bibnamefont {Wang}}, \bibinfo {author} {\bibfnamefont {W.}~\bibnamefont
  {Zhang}}, \bibinfo {author} {\bibfnamefont {M.~B.}\ \bibnamefont
  {Jungfleisch}}, \bibinfo {author} {\bibfnamefont {J.~E.}\ \bibnamefont
  {Pearson}}, \bibinfo {author} {\bibfnamefont {Y.}~\bibnamefont {Liu}},
  \bibinfo {author} {\bibfnamefont {J.}~\bibnamefont {Zang}}, \bibinfo {author}
  {\bibfnamefont {X.}~\bibnamefont {Cheng}}, \bibinfo {author} {\bibfnamefont
  {A.}~\bibnamefont {Petford-Long}}, \bibinfo {author} {\bibfnamefont
  {A.}~\bibnamefont {Hoffmann}}, \ and\ \bibinfo {author} {\bibfnamefont
  {S.~G.~E.}\ \bibnamefont {te~Velthuis}},\ }\href {\doibase
  10.1103/PhysRevB.99.104402} {\bibfield  {journal} {\bibinfo  {journal}
  {Physical Review B}\ }\textbf {\bibinfo {volume} {99}},\ \bibinfo {pages}
  {104402} (\bibinfo {year} {2019})}\BibitemShut {NoStop}%
\bibitem [{\citenamefont {Khadka}\ \emph {et~al.}(2018)\citenamefont {Khadka},
  \citenamefont {Karayev},\ and\ \citenamefont {Huang}}]{Khadka2018}%
  \BibitemOpen
  \bibfield  {author} {\bibinfo {author} {\bibfnamefont {D.}~\bibnamefont
  {Khadka}}, \bibinfo {author} {\bibfnamefont {S.}~\bibnamefont {Karayev}}, \
  and\ \bibinfo {author} {\bibfnamefont {S.~X.}\ \bibnamefont {Huang}},\ }\href
  {\doibase 10.1063/1.5021090} {\bibfield  {journal} {\bibinfo  {journal}
  {Journal of Applied Physics}\ }\textbf {\bibinfo {volume} {123}},\ \bibinfo
  {pages} {123905} (\bibinfo {year} {2018})}\BibitemShut {NoStop}%
\bibitem [{\citenamefont {Liu}\ \emph {et~al.}(2020)\citenamefont {Liu},
  \citenamefont {Zhao}, \citenamefont {Liu}, \citenamefont {Song},
  \citenamefont {Zhao},\ and\ \citenamefont {Zhang}}]{Liu_2020}%
  \BibitemOpen
  \bibfield  {author} {\bibinfo {author} {\bibfnamefont {L.}~\bibnamefont
  {Liu}}, \bibinfo {author} {\bibfnamefont {X.}~\bibnamefont {Zhao}}, \bibinfo
  {author} {\bibfnamefont {W.}~\bibnamefont {Liu}}, \bibinfo {author}
  {\bibfnamefont {Y.}~\bibnamefont {Song}}, \bibinfo {author} {\bibfnamefont
  {X.}~\bibnamefont {Zhao}}, \ and\ \bibinfo {author} {\bibfnamefont
  {Z.}~\bibnamefont {Zhang}},\ }\href {\doibase 10.1039/d0nr02168g} {\bibfield
  {journal} {\bibinfo  {journal} {Nanoscale}\ }\textbf {\bibinfo {volume}
  {12}},\ \bibinfo {pages} {12444} (\bibinfo {year} {2020})}\BibitemShut
  {NoStop}%
\bibitem [{\citenamefont {Schlotter}\ \emph {et~al.}(2018)\citenamefont
  {Schlotter}, \citenamefont {Agrawal},\ and\ \citenamefont
  {Beach}}]{Schlotter2018}%
  \BibitemOpen
  \bibfield  {author} {\bibinfo {author} {\bibfnamefont {S.}~\bibnamefont
  {Schlotter}}, \bibinfo {author} {\bibfnamefont {P.}~\bibnamefont {Agrawal}},
  \ and\ \bibinfo {author} {\bibfnamefont {G.~S.~D.}\ \bibnamefont {Beach}},\
  }\href {\doibase 10.1063/1.5038353} {\bibfield  {journal} {\bibinfo
  {journal} {Applied Physics Letters}\ }\textbf {\bibinfo {volume} {113}},\
  \bibinfo {pages} {092402} (\bibinfo {year} {2018})}\BibitemShut {NoStop}%
\bibitem [{\citenamefont {Michaelson}(1977)}]{Michaelson1977}%
  \BibitemOpen
  \bibfield  {author} {\bibinfo {author} {\bibfnamefont {H.~B.}\ \bibnamefont
  {Michaelson}},\ }\href {\doibase 10.1063/1.323539} {\bibfield  {journal}
  {\bibinfo  {journal} {Journal of Applied Physics}\ }\textbf {\bibinfo
  {volume} {48}},\ \bibinfo {pages} {4729} (\bibinfo {year}
  {1977})}\BibitemShut {NoStop}%
\bibitem [{\citenamefont {Kai}(2000)}]{Kai_2000}%
  \BibitemOpen
  \bibfield  {author} {\bibinfo {author} {\bibfnamefont {T.}~\bibnamefont
  {Kai}},\ }\href {\doibase 10.1016/S0304-8853(00)00082-2} {\bibfield
  {journal} {\bibinfo  {journal} {Journal of Magnetism and Magnetic Materials}\
  }\textbf {\bibinfo {volume} {214}},\ \bibinfo {pages} {167} (\bibinfo {year}
  {2000})}\BibitemShut {NoStop}%
\bibitem [{\citenamefont {Zhang}\ \emph {et~al.}(2014)\citenamefont {Zhang},
  \citenamefont {Wu},\ and\ \citenamefont {Kuch}}]{Zhang_2014}%
  \BibitemOpen
  \bibfield  {author} {\bibinfo {author} {\bibfnamefont {B.}~\bibnamefont
  {Zhang}}, \bibinfo {author} {\bibfnamefont {C.-B.}\ \bibnamefont {Wu}}, \
  and\ \bibinfo {author} {\bibfnamefont {W.}~\bibnamefont {Kuch}},\ }\href
  {\doibase 10.1063/1.4884235} {\bibfield  {journal} {\bibinfo  {journal}
  {Journal of Applied Physics}\ }\textbf {\bibinfo {volume} {115}},\ \bibinfo
  {pages} {233915} (\bibinfo {year} {2014})}\BibitemShut {NoStop}%
\bibitem [{\citenamefont {Henry}\ and\ \citenamefont
  {Ounadjela}(1996)}]{Henry_1996}%
  \BibitemOpen
  \bibfield  {author} {\bibinfo {author} {\bibfnamefont {Y.}~\bibnamefont
  {Henry}}\ and\ \bibinfo {author} {\bibfnamefont {K.}~\bibnamefont
  {Ounadjela}},\ }\href {\doibase 10.1103/PhysRevLett.76.1944} {\bibfield
  {journal} {\bibinfo  {journal} {Physical Review Letters}\ }\textbf {\bibinfo
  {volume} {76}},\ \bibinfo {pages} {1944} (\bibinfo {year}
  {1996})}\BibitemShut {NoStop}%
\bibitem [{\citenamefont {Ishida}\ and\ \citenamefont
  {Sagawa}(2004)}]{Ishida_2004}%
  \BibitemOpen
  \bibfield  {author} {\bibinfo {author} {\bibfnamefont {I.}~\bibnamefont
  {Ishida}}\ and\ \bibinfo {author} {\bibfnamefont {A.}~\bibnamefont
  {Sagawa}},\ }\href {\doibase 10.1016/j.jmmm.2004.01.088} {\bibfield
  {journal} {\bibinfo  {journal} {Journal of Magnetism and Magnetic Materials}\
  }\textbf {\bibinfo {volume} {279}},\ \bibinfo {pages} {261} (\bibinfo {year}
  {2004})}\BibitemShut {NoStop}%
\bibitem [{\citenamefont {Uchiyama}\ \emph {et~al.}(1996)\citenamefont
  {Uchiyama}, \citenamefont {Ishida}, \citenamefont {Hamada}, \citenamefont
  {Hirota}, \citenamefont {Okada},\ and\ \citenamefont
  {Hayashi}}]{Uchiyama_1996}%
  \BibitemOpen
  \bibfield  {author} {\bibinfo {author} {\bibfnamefont {K.}~\bibnamefont
  {Uchiyama}}, \bibinfo {author} {\bibfnamefont {I.}~\bibnamefont {Ishida}},
  \bibinfo {author} {\bibfnamefont {K.}~\bibnamefont {Hamada}}, \bibinfo
  {author} {\bibfnamefont {E.}~\bibnamefont {Hirota}}, \bibinfo {author}
  {\bibfnamefont {A.}~\bibnamefont {Okada}}, \ and\ \bibinfo {author}
  {\bibfnamefont {T.}~\bibnamefont {Hayashi}},\ }\href {\doibase
  10.1016/0304-8853(95)00927-2} {\bibfield  {journal} {\bibinfo  {journal}
  {Journal of Magnetism and Magnetic Materials}\ }\textbf {\bibinfo {volume}
  {156}},\ \bibinfo {pages} {429} (\bibinfo {year} {1996})}\BibitemShut
  {NoStop}%
\bibitem [{\citenamefont {Ishida}\ and\ \citenamefont
  {Morita}(2003)}]{Ishida_2003}%
  \BibitemOpen
  \bibfield  {author} {\bibinfo {author} {\bibfnamefont {I.}~\bibnamefont
  {Ishida}}\ and\ \bibinfo {author} {\bibfnamefont {S.}~\bibnamefont
  {Morita}},\ }\href {\doibase 10.1016/S0304-8853(03)00354-8} {\bibfield
  {journal} {\bibinfo  {journal} {Journal of Magnetism and Magnetic Materials}\
  }\textbf {\bibinfo {volume} {267}},\ \bibinfo {pages} {204} (\bibinfo {year}
  {2003})}\BibitemShut {NoStop}%
\bibitem [{\citenamefont {Nishimura}\ \emph {et~al.}(2021)\citenamefont
  {Nishimura}, \citenamefont {Kim}, \citenamefont {Kim}, \citenamefont {Nam},
  \citenamefont {Park}, \citenamefont {Kim}, \citenamefont {Shiota},
  \citenamefont {You}, \citenamefont {Min}, \citenamefont {Choe},\ and\
  \citenamefont {Ono}}]{Nishimura_2021}%
  \BibitemOpen
  \bibfield  {author} {\bibinfo {author} {\bibfnamefont {T.}~\bibnamefont
  {Nishimura}}, \bibinfo {author} {\bibfnamefont {D.-Y.}\ \bibnamefont {Kim}},
  \bibinfo {author} {\bibfnamefont {D.-H.}\ \bibnamefont {Kim}}, \bibinfo
  {author} {\bibfnamefont {Y.-S.}\ \bibnamefont {Nam}}, \bibinfo {author}
  {\bibfnamefont {Y.-K.}\ \bibnamefont {Park}}, \bibinfo {author}
  {\bibfnamefont {N.-H.}\ \bibnamefont {Kim}}, \bibinfo {author} {\bibfnamefont
  {Y.}~\bibnamefont {Shiota}}, \bibinfo {author} {\bibfnamefont {C.-Y.}\
  \bibnamefont {You}}, \bibinfo {author} {\bibfnamefont {B.-C.}\ \bibnamefont
  {Min}}, \bibinfo {author} {\bibfnamefont {S.-B.}\ \bibnamefont {Choe}}, \
  and\ \bibinfo {author} {\bibfnamefont {T.}~\bibnamefont {Ono}},\ }\href
  {\doibase 10.1103/PhysRevB.103.104409} {\bibfield  {journal} {\bibinfo
  {journal} {Physical Review B}\ }\textbf {\bibinfo {volume} {103}},\ \bibinfo
  {pages} {104409} (\bibinfo {year} {2021})}\BibitemShut {NoStop}%
\bibitem [{\citenamefont {Nishimura}\ \emph {et~al.}(2020)\citenamefont
  {Nishimura}, \citenamefont {Haruta}, \citenamefont {Kim}, \citenamefont
  {Shiota}, \citenamefont {Iwaki}, \citenamefont {Kan}, \citenamefont
  {Moriyama}, \citenamefont {Kurata},\ and\ \citenamefont
  {Ono}}]{Nishimura2020}%
  \BibitemOpen
  \bibfield  {author} {\bibinfo {author} {\bibfnamefont {T.}~\bibnamefont
  {Nishimura}}, \bibinfo {author} {\bibfnamefont {M.}~\bibnamefont {Haruta}},
  \bibinfo {author} {\bibfnamefont {D.-H.}\ \bibnamefont {Kim}}, \bibinfo
  {author} {\bibfnamefont {Y.}~\bibnamefont {Shiota}}, \bibinfo {author}
  {\bibfnamefont {H.}~\bibnamefont {Iwaki}}, \bibinfo {author} {\bibfnamefont
  {D.}~\bibnamefont {Kan}}, \bibinfo {author} {\bibfnamefont {T.}~\bibnamefont
  {Moriyama}}, \bibinfo {author} {\bibfnamefont {H.}~\bibnamefont {Kurata}}, \
  and\ \bibinfo {author} {\bibfnamefont {T.}~\bibnamefont {Ono}},\ }\href
  {\doibase 10.3379/msjmag.2001R002} {\bibfield  {journal} {\bibinfo  {journal}
  {Journal of the Magnetics Society of Japan}\ }\textbf {\bibinfo {volume}
  {44}},\ \bibinfo {pages} {9} (\bibinfo {year} {2020})}\BibitemShut {NoStop}%
\bibitem [{\citenamefont {Su}\ \emph {et~al.}(2009)\citenamefont {Su},
  \citenamefont {Lo}, \citenamefont {van Lierop}, \citenamefont {Lin},\ and\
  \citenamefont {Ouyang}}]{Su_2009}%
  \BibitemOpen
  \bibfield  {author} {\bibinfo {author} {\bibfnamefont {C.-H.}\ \bibnamefont
  {Su}}, \bibinfo {author} {\bibfnamefont {S.-C.}\ \bibnamefont {Lo}}, \bibinfo
  {author} {\bibfnamefont {J.}~\bibnamefont {van Lierop}}, \bibinfo {author}
  {\bibfnamefont {K.-W.}\ \bibnamefont {Lin}}, \ and\ \bibinfo {author}
  {\bibfnamefont {H.}~\bibnamefont {Ouyang}},\ }\href {\doibase
  10.1063/1.3070639} {\bibfield  {journal} {\bibinfo  {journal} {Journal of
  Applied Physics}\ }\textbf {\bibinfo {volume} {105}},\ \bibinfo {pages}
  {07C316} (\bibinfo {year} {2009})}\BibitemShut {NoStop}%
\bibitem [{\citenamefont {Bandiera}\ \emph {et~al.}(2012)\citenamefont
  {Bandiera}, \citenamefont {Sousa}, \citenamefont {Rodmacq},\ and\
  \citenamefont {Dieny}}]{Bandiera_2012}%
  \BibitemOpen
  \bibfield  {author} {\bibinfo {author} {\bibfnamefont {S.}~\bibnamefont
  {Bandiera}}, \bibinfo {author} {\bibfnamefont {R.~C.}\ \bibnamefont {Sousa}},
  \bibinfo {author} {\bibfnamefont {B.}~\bibnamefont {Rodmacq}}, \ and\
  \bibinfo {author} {\bibfnamefont {B.}~\bibnamefont {Dieny}},\ }\href
  {\doibase 10.1063/1.3701585} {\bibfield  {journal} {\bibinfo  {journal}
  {Applied Physics Letters}\ }\textbf {\bibinfo {volume} {100}},\ \bibinfo
  {pages} {142410} (\bibinfo {year} {2012})}\BibitemShut {NoStop}%
\bibitem [{\citenamefont {Wang}\ \emph {et~al.}(2020)\citenamefont {Wang},
  \citenamefont {Wei}, \citenamefont {He}, \citenamefont {Liu}, \citenamefont
  {Huang}, \citenamefont {Liu}, \citenamefont {Wang},\ and\ \citenamefont
  {Han}}]{Wang_2020}%
  \BibitemOpen
  \bibfield  {author} {\bibinfo {author} {\bibfnamefont {X.}~\bibnamefont
  {Wang}}, \bibinfo {author} {\bibfnamefont {Y.}~\bibnamefont {Wei}}, \bibinfo
  {author} {\bibfnamefont {K.}~\bibnamefont {He}}, \bibinfo {author}
  {\bibfnamefont {Y.}~\bibnamefont {Liu}}, \bibinfo {author} {\bibfnamefont
  {Y.}~\bibnamefont {Huang}}, \bibinfo {author} {\bibfnamefont
  {Q.}~\bibnamefont {Liu}}, \bibinfo {author} {\bibfnamefont {J.}~\bibnamefont
  {Wang}}, \ and\ \bibinfo {author} {\bibfnamefont {G.}~\bibnamefont {Han}},\
  }\href {\doibase 10.1088/1361-6463/ab78d7} {\bibfield  {journal} {\bibinfo
  {journal} {Journal of Physics D: Applied Physics}\ }\textbf {\bibinfo
  {volume} {53}},\ \bibinfo {pages} {215001} (\bibinfo {year}
  {2020})}\BibitemShut {NoStop}%
\bibitem [{\citenamefont {Lin}\ \emph {et~al.}(2018)\citenamefont {Lin},
  \citenamefont {Liu}, \citenamefont {Poellath}, \citenamefont {Zhang},
  \citenamefont {Ji}, \citenamefont {Lei}, \citenamefont {Yun}, \citenamefont
  {Xi}, \citenamefont {Yang}, \citenamefont {Xing}, \citenamefont {Wang},
  \citenamefont {Sun}, \citenamefont {Wu}, \citenamefont {Yin}, \citenamefont
  {Wang}, \citenamefont {Shen}, \citenamefont {Zweck}, \citenamefont {Back},
  \citenamefont {Zhang},\ and\ \citenamefont {Zhao}}]{Lin_2018}%
  \BibitemOpen
  \bibfield  {author} {\bibinfo {author} {\bibfnamefont {T.}~\bibnamefont
  {Lin}}, \bibinfo {author} {\bibfnamefont {H.}~\bibnamefont {Liu}}, \bibinfo
  {author} {\bibfnamefont {S.}~\bibnamefont {Poellath}}, \bibinfo {author}
  {\bibfnamefont {Y.}~\bibnamefont {Zhang}}, \bibinfo {author} {\bibfnamefont
  {B.}~\bibnamefont {Ji}}, \bibinfo {author} {\bibfnamefont {N.}~\bibnamefont
  {Lei}}, \bibinfo {author} {\bibfnamefont {J.~J.}\ \bibnamefont {Yun}},
  \bibinfo {author} {\bibfnamefont {L.}~\bibnamefont {Xi}}, \bibinfo {author}
  {\bibfnamefont {D.~Z.}\ \bibnamefont {Yang}}, \bibinfo {author}
  {\bibfnamefont {T.}~\bibnamefont {Xing}}, \bibinfo {author} {\bibfnamefont
  {Z.~L.}\ \bibnamefont {Wang}}, \bibinfo {author} {\bibfnamefont
  {L.}~\bibnamefont {Sun}}, \bibinfo {author} {\bibfnamefont {Y.~Z.}\
  \bibnamefont {Wu}}, \bibinfo {author} {\bibfnamefont {L.~F.}\ \bibnamefont
  {Yin}}, \bibinfo {author} {\bibfnamefont {W.~B.}\ \bibnamefont {Wang}},
  \bibinfo {author} {\bibfnamefont {J.}~\bibnamefont {Shen}}, \bibinfo {author}
  {\bibfnamefont {J.}~\bibnamefont {Zweck}}, \bibinfo {author} {\bibfnamefont
  {C.~H.}\ \bibnamefont {Back}}, \bibinfo {author} {\bibfnamefont {Y.~G.}\
  \bibnamefont {Zhang}}, \ and\ \bibinfo {author} {\bibfnamefont {W.~S.}\
  \bibnamefont {Zhao}},\ }\href {\doibase 10.1103/PhysRevB.98.174425}
  {\bibfield  {journal} {\bibinfo  {journal} {Physical Review B}\ }\textbf
  {\bibinfo {volume} {98}},\ \bibinfo {pages} {174425} (\bibinfo {year}
  {2018})}\BibitemShut {NoStop}%
\bibitem [{\citenamefont {Soumyanarayanan}\ \emph {et~al.}(2017)\citenamefont
  {Soumyanarayanan}, \citenamefont {Raju}, \citenamefont {Oyarce},
  \citenamefont {Tan}, \citenamefont {Im}, \citenamefont {Petrovi{\'{c}}},
  \citenamefont {Ho}, \citenamefont {Khoo}, \citenamefont {Tran}, \citenamefont
  {Gan}, \citenamefont {Ernult},\ and\ \citenamefont
  {Panagopoulos}}]{Soumyanarayanan_2017}%
  \BibitemOpen
  \bibfield  {author} {\bibinfo {author} {\bibfnamefont {A.}~\bibnamefont
  {Soumyanarayanan}}, \bibinfo {author} {\bibfnamefont {M.}~\bibnamefont
  {Raju}}, \bibinfo {author} {\bibfnamefont {A.~L.~G.}\ \bibnamefont {Oyarce}},
  \bibinfo {author} {\bibfnamefont {A.~K.~C.}\ \bibnamefont {Tan}}, \bibinfo
  {author} {\bibfnamefont {M.-Y.}\ \bibnamefont {Im}}, \bibinfo {author}
  {\bibfnamefont {A.~P.}\ \bibnamefont {Petrovi{\'{c}}}}, \bibinfo {author}
  {\bibfnamefont {P.}~\bibnamefont {Ho}}, \bibinfo {author} {\bibfnamefont
  {K.~H.}\ \bibnamefont {Khoo}}, \bibinfo {author} {\bibfnamefont
  {M.}~\bibnamefont {Tran}}, \bibinfo {author} {\bibfnamefont {C.~K.}\
  \bibnamefont {Gan}}, \bibinfo {author} {\bibfnamefont {F.}~\bibnamefont
  {Ernult}}, \ and\ \bibinfo {author} {\bibfnamefont {C.}~\bibnamefont
  {Panagopoulos}},\ }\href {\doibase 10.1038/nmat4934} {\bibfield  {journal}
  {\bibinfo  {journal} {Nature Materials}\ }\textbf {\bibinfo {volume} {16}},\
  \bibinfo {pages} {898} (\bibinfo {year} {2017})}\BibitemShut {NoStop}%
\bibitem [{\citenamefont {Men'shikov}\ \emph {et~al.}(1985)\citenamefont
  {Men'shikov}, \citenamefont {Takzei}, \citenamefont {Dorofeev}, \citenamefont
  {Kazantsev}, \citenamefont {Kostyshin},\ and\ \citenamefont
  {Sych}}]{Menshikov1985}%
  \BibitemOpen
  \bibfield  {author} {\bibinfo {author} {\bibfnamefont {A.~Z.}\ \bibnamefont
  {Men'shikov}}, \bibinfo {author} {\bibfnamefont {G.}~\bibnamefont {Takzei}},
  \bibinfo {author} {\bibfnamefont {Y.~A.}\ \bibnamefont {Dorofeev}}, \bibinfo
  {author} {\bibfnamefont {V.~A.}\ \bibnamefont {Kazantsev}}, \bibinfo {author}
  {\bibfnamefont {A.~K.}\ \bibnamefont {Kostyshin}}, \ and\ \bibinfo {author}
  {\bibfnamefont {I.~I.}\ \bibnamefont {Sych}},\ }\href@noop {} {\bibfield
  {journal} {\bibinfo  {journal} {Zh. Eksp. Teor. Fiz.}\ }\textbf {\bibinfo
  {volume} {89}},\ \bibinfo {pages} {1269} (\bibinfo {year}
  {1985})}\BibitemShut {NoStop}%
\bibitem [{\citenamefont {Waring}\ \emph {et~al.}(2020)\citenamefont {Waring},
  \citenamefont {Johansson}, \citenamefont {Vera-Marun},\ and\ \citenamefont
  {Thomson}}]{Waring_2020}%
  \BibitemOpen
  \bibfield  {author} {\bibinfo {author} {\bibfnamefont {H.~J.}\ \bibnamefont
  {Waring}}, \bibinfo {author} {\bibfnamefont {N.~A.~B.}\ \bibnamefont
  {Johansson}}, \bibinfo {author} {\bibfnamefont {I.~J.}\ \bibnamefont
  {Vera-Marun}}, \ and\ \bibinfo {author} {\bibfnamefont {T.}~\bibnamefont
  {Thomson}},\ }\href {\doibase 10.1103/PhysRevApplied.13.034035} {\bibfield
  {journal} {\bibinfo  {journal} {Physical Review Applied}\ }\textbf {\bibinfo
  {volume} {13}},\ \bibinfo {pages} {034035} (\bibinfo {year}
  {2020})}\BibitemShut {NoStop}%
\bibitem [{\citenamefont {Carcia}(1988)}]{Carcia_1988}%
  \BibitemOpen
  \bibfield  {author} {\bibinfo {author} {\bibfnamefont {P.~F.}\ \bibnamefont
  {Carcia}},\ }\href {\doibase 10.1063/1.340404} {\bibfield  {journal}
  {\bibinfo  {journal} {Journal of Applied Physics}\ }\textbf {\bibinfo
  {volume} {63}},\ \bibinfo {pages} {5066} (\bibinfo {year}
  {1988})}\BibitemShut {NoStop}%
\bibitem [{\citenamefont {Zeper}\ \emph {et~al.}(1989)\citenamefont {Zeper},
  \citenamefont {Greidanus}, \citenamefont {Carcia},\ and\ \citenamefont
  {Fincher}}]{Zeper_1989}%
  \BibitemOpen
  \bibfield  {author} {\bibinfo {author} {\bibfnamefont {W.~B.}\ \bibnamefont
  {Zeper}}, \bibinfo {author} {\bibfnamefont {F.~J. A.~M.}\ \bibnamefont
  {Greidanus}}, \bibinfo {author} {\bibfnamefont {P.~F.}\ \bibnamefont
  {Carcia}}, \ and\ \bibinfo {author} {\bibfnamefont {C.~R.}\ \bibnamefont
  {Fincher}},\ }\href {\doibase 10.1063/1.343189} {\bibfield  {journal}
  {\bibinfo  {journal} {Journal of Applied Physics}\ }\textbf {\bibinfo
  {volume} {65}},\ \bibinfo {pages} {4971} (\bibinfo {year}
  {1989})}\BibitemShut {NoStop}%
\bibitem [{\citenamefont {Denker}\ \emph {et~al.}(2020)\citenamefont {Denker},
  \citenamefont {Nielsen}, \citenamefont {Lage}, \citenamefont {Römer-Stumm},
  \citenamefont {Heyen}, \citenamefont {Junk}, \citenamefont {Walowski},
  \citenamefont {Waldorf}, \citenamefont {Münzenberg},\ and\ \citenamefont
  {McCord}}]{Denker_2020}%
  \BibitemOpen
  \bibfield  {author} {\bibinfo {author} {\bibfnamefont {C.}~\bibnamefont
  {Denker}}, \bibinfo {author} {\bibfnamefont {S.}~\bibnamefont {Nielsen}},
  \bibinfo {author} {\bibfnamefont {E.}~\bibnamefont {Lage}}, \bibinfo {author}
  {\bibfnamefont {M.}~\bibnamefont {Römer-Stumm}}, \bibinfo {author}
  {\bibfnamefont {H.}~\bibnamefont {Heyen}}, \bibinfo {author} {\bibfnamefont
  {Y.}~\bibnamefont {Junk}}, \bibinfo {author} {\bibfnamefont {J.}~\bibnamefont
  {Walowski}}, \bibinfo {author} {\bibfnamefont {K.}~\bibnamefont {Waldorf}},
  \bibinfo {author} {\bibfnamefont {M.}~\bibnamefont {Münzenberg}}, \ and\
  \bibinfo {author} {\bibfnamefont {J.}~\bibnamefont {McCord}},\ }\href@noop {}
  {\  (\bibinfo {year} {2020})},\ \Eprint
  {http://arxiv.org/abs/https://arxiv.org/abs/2011.07336}
  {arXiv:https://arxiv.org/abs/2011.07336 [cond-mat.mes-hall]} \BibitemShut
  {NoStop}%
\bibitem [{\citenamefont {Thomas}\ \emph {et~al.}(2012)\citenamefont {Thomas},
  \citenamefont {Pookat}, \citenamefont {Nair}, \citenamefont {Daniel},
  \citenamefont {Dymerska}, \citenamefont {Liebig}, \citenamefont {Al-Harthi},
  \citenamefont {Ramanujan}, \citenamefont {Anantharaman}, \citenamefont
  {Fidler},\ and\ \citenamefont {Albrecht}}]{Thomas_2012}%
  \BibitemOpen
  \bibfield  {author} {\bibinfo {author} {\bibfnamefont {S.}~\bibnamefont
  {Thomas}}, \bibinfo {author} {\bibfnamefont {G.}~\bibnamefont {Pookat}},
  \bibinfo {author} {\bibfnamefont {S.~S.}\ \bibnamefont {Nair}}, \bibinfo
  {author} {\bibfnamefont {M.}~\bibnamefont {Daniel}}, \bibinfo {author}
  {\bibfnamefont {B.}~\bibnamefont {Dymerska}}, \bibinfo {author}
  {\bibfnamefont {A.}~\bibnamefont {Liebig}}, \bibinfo {author} {\bibfnamefont
  {S.~H.}\ \bibnamefont {Al-Harthi}}, \bibinfo {author} {\bibfnamefont {R.~V.}\
  \bibnamefont {Ramanujan}}, \bibinfo {author} {\bibfnamefont {M.~R.}\
  \bibnamefont {Anantharaman}}, \bibinfo {author} {\bibfnamefont
  {J.}~\bibnamefont {Fidler}}, \ and\ \bibinfo {author} {\bibfnamefont
  {M.}~\bibnamefont {Albrecht}},\ }\href {\doibase
  10.1088/0953-8984/24/25/256004} {\bibfield  {journal} {\bibinfo  {journal}
  {Journal of Physics: Condensed Matter}\ }\textbf {\bibinfo {volume} {24}},\
  \bibinfo {pages} {256004} (\bibinfo {year} {2012})}\BibitemShut {NoStop}%
\bibitem [{\citenamefont {Luo}\ \emph {et~al.}(2015)\citenamefont {Luo},
  \citenamefont {Fu}, \citenamefont {Zhang}, \citenamefont {Yuan},
  \citenamefont {Zhai}, \citenamefont {Dong},\ and\ \citenamefont
  {Zhai}}]{Luo_2015}%
  \BibitemOpen
  \bibfield  {author} {\bibinfo {author} {\bibfnamefont {C.}~\bibnamefont
  {Luo}}, \bibinfo {author} {\bibfnamefont {Y.}~\bibnamefont {Fu}}, \bibinfo
  {author} {\bibfnamefont {D.}~\bibnamefont {Zhang}}, \bibinfo {author}
  {\bibfnamefont {S.}~\bibnamefont {Yuan}}, \bibinfo {author} {\bibfnamefont
  {Y.}~\bibnamefont {Zhai}}, \bibinfo {author} {\bibfnamefont {S.}~\bibnamefont
  {Dong}}, \ and\ \bibinfo {author} {\bibfnamefont {H.}~\bibnamefont {Zhai}},\
  }\href {\doibase 10.1016/j.jmmm.2014.09.014} {\bibfield  {journal} {\bibinfo
  {journal} {Journal of Magnetism and Magnetic Materials}\ }\textbf {\bibinfo
  {volume} {374}},\ \bibinfo {pages} {711} (\bibinfo {year}
  {2015})}\BibitemShut {NoStop}%
\bibitem [{\citenamefont {Schäfer}(2007)}]{Schaefer_2007}%
  \BibitemOpen
  \bibfield  {author} {\bibinfo {author} {\bibfnamefont {R.}~\bibnamefont
  {Schäfer}},\ }\href {\doibase 10.1002/9780470022184.hmm310} {\bibfield
  {journal} {\bibinfo  {journal} {Handbook of Magnetism and Advanced Magnetic
  Materials}\ } (\bibinfo {year} {2007}),\
  10.1002/9780470022184.hmm310}\BibitemShut {NoStop}%
\bibitem [{\citenamefont {Woodward}\ \emph {et~al.}(2003)\citenamefont
  {Woodward}, \citenamefont {Lance}, \citenamefont {Street},\ and\
  \citenamefont {Stamps}}]{Woodward_2003}%
  \BibitemOpen
  \bibfield  {author} {\bibinfo {author} {\bibfnamefont {R.~C.}\ \bibnamefont
  {Woodward}}, \bibinfo {author} {\bibfnamefont {A.~M.}\ \bibnamefont {Lance}},
  \bibinfo {author} {\bibfnamefont {R.}~\bibnamefont {Street}}, \ and\ \bibinfo
  {author} {\bibfnamefont {R.~L.}\ \bibnamefont {Stamps}},\ }\href {\doibase
  10.1063/1.1557654} {\bibfield  {journal} {\bibinfo  {journal} {Journal of
  Applied Physics}\ }\textbf {\bibinfo {volume} {93}},\ \bibinfo {pages} {6567}
  (\bibinfo {year} {2003})}\BibitemShut {NoStop}%
\bibitem [{\citenamefont {Bandiera}\ \emph {et~al.}(2013)\citenamefont
  {Bandiera}, \citenamefont {Sousa}, \citenamefont {Rodmacq}, \citenamefont
  {Lechevallier},\ and\ \citenamefont {Dieny}}]{Bandiera_2013}%
  \BibitemOpen
  \bibfield  {author} {\bibinfo {author} {\bibfnamefont {S.}~\bibnamefont
  {Bandiera}}, \bibinfo {author} {\bibfnamefont {R.~C.}\ \bibnamefont {Sousa}},
  \bibinfo {author} {\bibfnamefont {B.}~\bibnamefont {Rodmacq}}, \bibinfo
  {author} {\bibfnamefont {L.}~\bibnamefont {Lechevallier}}, \ and\ \bibinfo
  {author} {\bibfnamefont {B.}~\bibnamefont {Dieny}},\ }\href {\doibase
  10.1088/0022-3727/46/48/485003} {\bibfield  {journal} {\bibinfo  {journal}
  {Journal of Physics D: Applied Physics}\ }\textbf {\bibinfo {volume} {46}},\
  \bibinfo {pages} {485003} (\bibinfo {year} {2013})}\BibitemShut {NoStop}%
\bibitem [{\citenamefont {Morrow}\ \emph {et~al.}(2015)\citenamefont {Morrow},
  \citenamefont {Pearton},\ and\ \citenamefont {Ren}}]{Morrow_2015}%
  \BibitemOpen
  \bibfield  {author} {\bibinfo {author} {\bibfnamefont {W.~K.}\ \bibnamefont
  {Morrow}}, \bibinfo {author} {\bibfnamefont {S.~J.}\ \bibnamefont {Pearton}},
  \ and\ \bibinfo {author} {\bibfnamefont {F.}~\bibnamefont {Ren}},\ }\href
  {\doibase 10.1002/smll.201501120} {\bibfield  {journal} {\bibinfo  {journal}
  {Small}\ }\textbf {\bibinfo {volume} {12}},\ \bibinfo {pages} {120} (\bibinfo
  {year} {2015})}\BibitemShut {NoStop}%
\bibitem [{\citenamefont {Hong}\ \emph {et~al.}(2014)\citenamefont {Hong},
  \citenamefont {Lee}, \citenamefont {Lee}, \citenamefont {Han}, \citenamefont
  {Mahata}, \citenamefont {Yeon}, \citenamefont {Koo}, \citenamefont {Kim},
  \citenamefont {Nam}, \citenamefont {Byun}, \citenamefont {Min}, \citenamefont
  {Kim}, \citenamefont {Kim}, \citenamefont {Joo},\ and\ \citenamefont
  {Lee}}]{Hong_2014}%
  \BibitemOpen
  \bibfield  {author} {\bibinfo {author} {\bibfnamefont {J.}~\bibnamefont
  {Hong}}, \bibinfo {author} {\bibfnamefont {S.}~\bibnamefont {Lee}}, \bibinfo
  {author} {\bibfnamefont {S.}~\bibnamefont {Lee}}, \bibinfo {author}
  {\bibfnamefont {H.}~\bibnamefont {Han}}, \bibinfo {author} {\bibfnamefont
  {C.}~\bibnamefont {Mahata}}, \bibinfo {author} {\bibfnamefont {H.-W.}\
  \bibnamefont {Yeon}}, \bibinfo {author} {\bibfnamefont {B.}~\bibnamefont
  {Koo}}, \bibinfo {author} {\bibfnamefont {S.-I.}\ \bibnamefont {Kim}},
  \bibinfo {author} {\bibfnamefont {T.}~\bibnamefont {Nam}}, \bibinfo {author}
  {\bibfnamefont {K.}~\bibnamefont {Byun}}, \bibinfo {author} {\bibfnamefont
  {B.-W.}\ \bibnamefont {Min}}, \bibinfo {author} {\bibfnamefont {Y.-W.}\
  \bibnamefont {Kim}}, \bibinfo {author} {\bibfnamefont {H.}~\bibnamefont
  {Kim}}, \bibinfo {author} {\bibfnamefont {Y.-C.}\ \bibnamefont {Joo}}, \ and\
  \bibinfo {author} {\bibfnamefont {T.}~\bibnamefont {Lee}},\ }\href {\doibase
  10.1039/C3NR06771H} {\bibfield  {journal} {\bibinfo  {journal} {Nanoscale}\
  }\textbf {\bibinfo {volume} {6}},\ \bibinfo {pages} {7503} (\bibinfo {year}
  {2014})}\BibitemShut {NoStop}%
\bibitem [{\citenamefont {Polisetty}\ \emph {et~al.}(2008)\citenamefont
  {Polisetty}, \citenamefont {Scheffler}, \citenamefont {Sahoo}, \citenamefont
  {Wang}, \citenamefont {Mukherjee}, \citenamefont {He},\ and\ \citenamefont
  {Binek}}]{Polisetty_2008}%
  \BibitemOpen
  \bibfield  {author} {\bibinfo {author} {\bibfnamefont {S.}~\bibnamefont
  {Polisetty}}, \bibinfo {author} {\bibfnamefont {J.}~\bibnamefont
  {Scheffler}}, \bibinfo {author} {\bibfnamefont {S.}~\bibnamefont {Sahoo}},
  \bibinfo {author} {\bibfnamefont {Y.}~\bibnamefont {Wang}}, \bibinfo {author}
  {\bibfnamefont {T.}~\bibnamefont {Mukherjee}}, \bibinfo {author}
  {\bibfnamefont {X.}~\bibnamefont {He}}, \ and\ \bibinfo {author}
  {\bibfnamefont {C.}~\bibnamefont {Binek}},\ }\href {\doibase
  10.1063/1.2932445} {\bibfield  {journal} {\bibinfo  {journal} {Review of
  Scientific Instruments}\ }\textbf {\bibinfo {volume} {79}},\ \bibinfo {pages}
  {055107} (\bibinfo {year} {2008})}\BibitemShut {NoStop}%
\bibitem [{\citenamefont {Beaujour}\ \emph {et~al.}(2006)\citenamefont
  {Beaujour}, \citenamefont {Chen}, \citenamefont {Kent},\ and\ \citenamefont
  {Sun}}]{Beaujour_2006}%
  \BibitemOpen
  \bibfield  {author} {\bibinfo {author} {\bibfnamefont {J.-M.~L.}\
  \bibnamefont {Beaujour}}, \bibinfo {author} {\bibfnamefont {W.}~\bibnamefont
  {Chen}}, \bibinfo {author} {\bibfnamefont {A.~D.}\ \bibnamefont {Kent}}, \
  and\ \bibinfo {author} {\bibfnamefont {J.~Z.}\ \bibnamefont {Sun}},\ }\href
  {\doibase 10.1063/1.2151832} {\bibfield  {journal} {\bibinfo  {journal}
  {Journal of Applied Physics}\ }\textbf {\bibinfo {volume} {99}},\ \bibinfo
  {pages} {08N503} (\bibinfo {year} {2006})}\BibitemShut {NoStop}%
\end{thebibliography}%


\clearpage

\end{document}